\documentclass[journal,11pt,draftcls,onecolumn]{IEEEtran}%
\usepackage{amsfonts}
\usepackage{amsmath}
\usepackage{amssymb}
\usepackage{graphicx}
\usepackage{cite}%
\providecommand{\U}[1]{\protect\rule{.1in}{.1in}}
\newtheorem{theorem}{Theorem}

\newtheorem{definition}{Definition}
\newtheorem{example}{Example}

\newtheorem{proposition}{Proposition}
\newtheorem{remark}{Remark}

\begin{document}

\title{Utility-Privacy Tradeoff in Databases: An Information-theoretic Approach }
\author{Lalitha Sankar,~\IEEEmembership{Member,~IEEE,} S. Raj Rajagopalan,
and~H.\ Vincent Poor,~\IEEEmembership{Fellow,~IEEE}\thanks{L. Sankar is with
the Department of Electrical, Computer, and Energy Engineering at Arizona
State University. S. Raj Rajagopalan is with Honeywell ACS Labs. H. V. Poor is
with the Department of Electrical\ Engineering at Princeton University. Part
of this work was done when L.\ Sankar was at Princeton University and S.
Rajagopalan was with HP\ Labs. email:
\{lalithasankar@asu.edu,siva.rajagopalan@honeywell.com,poor@princeton.edu\}.}%
\thanks{This research was supported in part by the NSF\ under Grants
CNS-09-05398 and CCF-10-16671 and the AFOSR\ under Grant FA9550-09-1-0643. The
material in this paper were presented in part at the International Symposium
on Information Theory 2010 and the Allerton\ Conference on Control, Computing
and Communications, Monticello, IL, 2010.}}
\maketitle

\begin{abstract}
Ensuring the usefulness of electronic data sources while providing necessary
privacy guarantees is an important unsolved problem. This problem drives the
need for an analytical framework that can quantify the privacy of personally
identifiable information while still providing a quantifable benefit (utility)
to multiple legitimate information consumers. This paper presents an
information-theoretic framework that promises an analytical model guaranteeing
tight bounds of how much utility is possible for a given level of privacy and
vice-versa. Specific contributions include: i) stochastic data models for both
categorical and numerical data; ii) utility-privacy tradeoff regions and the
encoding (sanization) schemes achieving them for both classes and their
practical relevance; and iii)\ modeling of prior knowledge at the user and/or
data source and optimal encoding schemes for both cases.

\end{abstract}


\section{Introduction}

\begin{figure}[ptb]%
\centering
\includegraphics[
height=1.9735in,
width=3.1462in
]%
{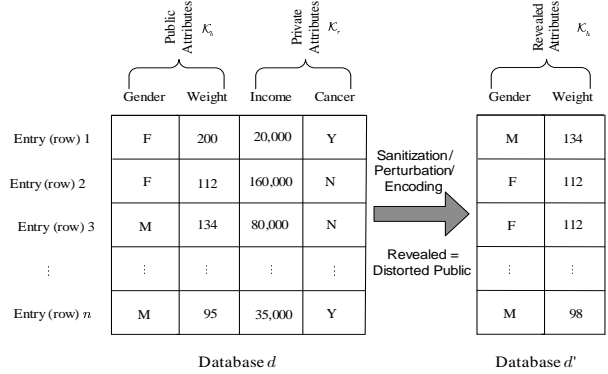}%
\caption{An example database with public and private attributes and its
sanitized version.}%
\label{FigDB}%
\end{figure}

Just as information technology and electronic communications have been rapidly
applied to almost every sphere of human activity, including commerce, medicine
and social networking, the risk of accidental or intentional disclosure of
sensitive private information has increased. The concomitant creation of large
centralized searchable data repositories and deployment of applications that
use them has made \textquotedblleft leakage\textquotedblright\ of private
information such as medical data, credit card information, power consumption
data, etc. highly probable and thus an important and urgent societal problem.
Unlike the secrecy problem, in the \emph{privacy} problem, disclosing data
provides informational utility while enabling possible loss of privacy at the
same time. Thus, as shown in\ Fig. \ref{FigDB}, in the course of a legitimate
transaction, a user learns some public information (e.g. gender and weight),
which is allowed and needs to be supported for the transaction to be
meaningful, and at the same time he can also learn/infer private information
(e.g., cancer and income), which needs to be prevented (or minimized). Thus,
every user is (potentially)\ also an adversary.

The problem of privacy and information leakage has been studied for several
decades by multiple research communities; information-theoretic approaches to
the problem are few and far in between and have primarily focused on using
information-theoretic metrics. However, a rigorous information-theoretic
treatment of the utility-privacy (U-P) tradeoff problem remains open and the
following questions are yet to be addressed: (i) the statistical assumptions
on the data that allow information-theoretic analysis, (ii) the capability of
revealing different levels of private information to different users, and
(iii) modeling of and accounting for prior knowledge. In this work, we seek to
apply information theoretic tools to address the open question of an
analytical characterization that provides a tight U-P tradeoff. If one views
public and private attributes of data in a repository as random variables with
a joint probability distribution, a private attribute in a database remains
private to the extent that revealing public attributes releases no additional
information about it -- in other words, minimizing the risk of privacy loss
implies that \textit{the conditional entropy of the private attribute should
be as high as possible after the disclosure}. Thus, in\ Fig. \ref{FigDB},
keeping the cancer attribute private would mean that, given knowledge of the
public attributes of gender and weight, the predictability of the cancer
attribute should remain unchanged. To achieve this, the gender attribute in
Entry 1 has been \textquotedblleft sanitized.\textquotedblright%

The utility of a data source lies in its ability to disclose data and privacy
considerations have the potential to hurt utility. Indeed, utility and privacy
are competing goals in this context. For example, in \ Fig. \ref{FigDB} one
could sanitize all or most of the entries in the gender attribute to `M' to
obtain more privacy but that could reduce the usefulness of the published data
significantly. Any approach that considers only the privacy aspect of
information disclosure while ignoring the resultant reduction in utility is
not likely to be practically viable. To make a reasoned tradeoff, we need to
know the maximum utility achievable for a given level of privacy and vice
versa, i.e.\ an analytical characterization of the set of all achievable U-P
tradeoff points. We show that this can be done using an elegant tool from
information theory called rate distortion theory: utility can be quantified
via fidelity which, in turn, is related (inversely) to \textit{distortion}.
Rate distortion has to be augmented with privacy constraints quantified via
\textit{equivocation}, which is related to entropy.

\textit{Our Contributions: }The central contribution of this work is a precise
quantification of the tradeoff between the privacy needs of the individuals
represented by the data and the utility of the \textit{sanitized} (published)
data for any data source using the theory of rate distortion with additional
privacy constraints. Utility is quantified (inversely) via \textit{distortion
}(accuracy), and privacy via \textit{equivocation} (entropy).

We expose for the first time an essential dimension of information disclosure
via an additional constraint on the disclosure rate, a measure of the
precision of the sanitized data. Any controlled disclosure of public data
needs to specify the accuracy and precision of the disclosure; while the two
can be conflated using additive noise for numerical data, additive noise is
not an option for categorical data (social security numbers, postal codes,
disease status, etc.) and thus output precision becomes important to specify.
For example, in\ \ Fig. \ref{FigDB}, the weight attribute is a numeric field
that could either be distorted with random additive noise or truncated (or
quantized) into ranges such as 90-100, 100-110, etc. The use of the digits of
the social security number to identify and protect the privacy of students in
grade sheets is a familiar non-numeric example. Sanitization (of the full SSN)
is achieved by heuristically reducing precision to typically the last four
digits. A theoretical framework that formally specifies the output precision
necessary and sufficient to achieve the optimal U-P tradeoff would be desirable.

In \cite{Yamamoto} the rate-distortion-equivocation (RDE) tradeoff for a
simple source model was presented. We translate this formalism to the U-P
problem and develop a \textit{framework that allows us to model generic data
sources}, including multi-dimensional databases and data streams
\cite{LS_SG4}, develop abstract utility and privacy metrics, and quantify the
fundamental U-P tradeoff bounds. We then present a \textit{sanitization scheme
that achieves the U-P tradeoff region} and demonstrate the application of this
scheme for both numerical and categorical examples. Noting that correlation
available to the user/adversary can be internal (i.e.\ between variables
within a database) or external (with variables that are outside the database
but accessible to the user/adversary), \cite{Sweeney,Dwork_DP,Shmatikov} have
shown that external knowledge can be very powerful in the privacy context. We
address this challenge in our framework via a \textit{model for side
information}. Our theorem in this context reported previously in \cite{SRV3}
is presented with the full proof here.

Finally, we demonstrate our framework with two crucial and practically
relevant examples: categorical and numerical databases. Our examples
demonstrate two fundamental aspects of our framework: (i) how statistical
models for the data and U-P metrics reveals the appropriate distortion and
suppression of data to achieve both privacy and utility guarantees; and (ii)
how knowledge of source statistics enables determining the U-P optimal
sanitization mechanism, and therefore, the largest U-P tradeoff region.

The paper is organized as follows. In Section \ref{Sec_II} we briefly
summarize the state of the art in database privacy research. In Section
\ref{Sec_MotBack}, we motivate the need for an information-theoretic analysis
and present the intuition behind our analytical framework. In Section
\ref{Sec_Mod}, we present an abstract model and metrics for structured data
sources such as databases. We develop our primary analytical framework in
Section \ref{Sec_UPT} and illustrate our results in Section \ref{Sec_Illus}.
We close with concluding remarks in Section \ref{Sec_V}.

\section{\label{Sec_II}Related Work}

The problem of privacy in databases has a long and rich history dating back at
least to the 1970s, and space restrictions preclude any attempt to do full
justice to the different approaches that have been considered along the way.
We divide the existing work into two categories, heuristic and theoretical
techniques, and outline the major milestones from these categories for comparison.

The earliest attempts at systematic privacy were in the area of census data
publication where data was required to be made public but without leaking
individuals' information. A number of ad hoc techniques such as sub-sampling,
aggregation, and suppression were explored (e.g., \cite{Raghu,Dobra} and the
references therein). The first formal definition of privacy was $k$-anonymity
by Sweeney \cite{Sweeney}. However $k$-anonymity was found to be inadequate as
it only protects from identity disclosure but not attribute-based disclosure
and was extended with $t$-closeness \cite{Chawla02} and $l$-diversity
\cite{L-diversity}. All these techniques have proved to be non-universal as
they were only robust against limited adversaries. Heuristic techniques for
privacy in data mining have focused on using a mutual information-based
privacy metrics \cite{Ag_Ag}.

The first universal formalism was proposed in differential privacy (DP)
\cite{Dwork_DP} (see the survey in \cite{Dwork_ACM} for a detailed history of
the field). In this model, the privacy of an individual in a database is
defined as a bound on the ability of any adversary to accurately detect
whether that individual's data belongs to the database or not. They also show
that Laplacian distributed additive noise with appropriately chosen parameters
suffices to sanitize numerical data to achieve differential privacy. The
concept of DP is strictly stronger than our definition of privacy, which is
based on Shannon entropy. However, our model seems more intuitively accessible
and suited to many application domains where strict anonymity is not the
requirement. For example, in many wellness databases the presence of the
record of an individual is not a secret but that individual's disease status
is. Our sanitization approach applies to both numerical and categorical data
whereas DP, while being a very popular model for privacy, appears limited to
numerical data. Furthermore, the loss of utility from DP-based sanitization
can be significant \cite{GJagannathan}. There has been some work pointing out
the loss of utility due to privacy mechanisms for specific applications
\cite{data-publishing}.

More generally, a rigorous model for privacy-utility tradeoffs with a method
to achieve \emph{all} the optimal points has remained open and is the subject
of this paper. The use of information theoretic tools for privacy and related
problems is relatively sparse. \cite{Yamamoto} analyzed a simple two variable
model using rate distortion theory with equivocation constraints, which is the
prime motivation for this work. In addition, there has been recent work
comparing differential privacy guarantee with Renyi entropy \cite{Alvim} and
Shannon entropy \cite{Fawaz}.

\section{\label{Sec_MotBack}Motivation and Background}

The information-theoretic approach to database privacy involves two steps: the
first is the data modeling step and the second is deriving the mathematical
formalism for sanitization. Before we introduce our formal model and
abstractions, we first present an intuitive understanding and motivation for
our approaches below.

\subsection{Motivation: Statistical Model}

Our work is based on the observation that large datasets (including databases)
have a distributional basis; i.e., there exists an underlying (sometimes
implicit) statistical model for the data. Even in the case of data mining
where only one or a few instances of the dataset are ever available, the use
of correlations between attributes used an implicit distributional assumption
about the dataset. We explicitly model the data as being generated by a source
with a finite or infinite alphabet and a known distribution. Each row of the
database is a collection of correlated attributes (of an individual) that
belongs to the alphabet of the source and is generated according to the
probability of occurrence of that letter (of the alphabet).

Our statistical model for databases is also motivated by the fact that while
the attributes of an individual may be correlated (e.g.\ between the weight
and cancer attributes in\ Fig. \ref{FigDB}), the records of a large number of
individuals are generally independent or weakly correlated with each other. We
thus model the database as a collection of $n$ observations generated by a
memoryless source whose outputs are independent and identically distributed (i.i.d.).

Statistically, with a large number $n$ of i.i.d. samples collected from a
source, the data collected can be viewed as \textit{typical}, i.e., it follows
the strong law of large numbers (SLLN) \cite[Ch. 11]{CTbook}. The SLLN implies
that the absolute difference between the empirical distribution (obtained from
the observations) and the actual distribution of each letter of the source
alphabet decreases with $n$, i.e., the samples (letters from the source
alphabet) in the database will be represented proportional to their actual
probabilities. This implies that for all practical purposes the empirical
distribution obtained from a large dataset can be assumed to be the
statistical distribution of the idealized source for our model and the
approximation gets better as $n$ grows.

Our measures for utility and privacy capture this statistical model. In
particular, we quantify privacy using \textit{conditional entropy} where the
conditioning on the published (revealed) data captures the average uncertainty
about the source (specifically, the private attributes of the source)
post-sanitization. Our utility measure similarly is averaged over the source distribution.

Intuitively, privacy is about maintaining uncertainty about information that
is not explicitly disclosed. The common notion of a person being undetectable
in a group as in \cite{Sweeney} or an individual record remaining undetectable
in a dataset \cite{Dwork_DP} captures one flavor of such uncertainty. More
generally, the uncertainty about a piece of undisclosed information is related
to its information content. Our approach focuses on the information content of
every sample of the source and sanitizes it in proportion to its likelihood in
the database. This, in turn, ensures that low probability/high information
samples (outliers) are suppressed or heavily distorted whereas the high
probability (frequent flier) samples are distorted only slightly. Outlier
data, if released without sanitization, can leak a lot of information to the
adversary about those individuals (e.g.\ individuals older than a hundred
years); on the other hand, for individuals represented by high probability
samples either the adversary already has a lot of information about them or
they are sufficiently indistinct due to their high occurrence in the data,
thereby allowing smaller distortion.

As we show formally in the sequel, our approach and solution for categorical
databases captures a critical aspect of the privacy challenge, namely, in
suppressing the high information (low probability outlier samples) and
distorting all others (up to the desired utility/distortion level), the
database provides uncertainty (for that distortion level) for \emph{all}
samples of the data. Thus, our statistical privacy measure captures the
characteristics of the underlying data model.

It is crucial to note that distortion does not only imply distance-based
measures. The distortion measure can be chosen to preserve any desired
function, deterministic or probabilistic, of the attributes (e.g., aggregate
statistics). Our aim is to ensure that sensitive data is protected by
randomizing the public (non-sensitive) data in a rigorous and well-defined
manner such that: (a) it still preserves some measure of the original public
data\ (e.g., K-L divergence, Euclidean distance, Hamming distortion, etc.);
and (b) provides some measure of privacy for the sensitive data that can be
inferred from the revealed data. In this context, distortion is a term that
makes precise a measure of change between the original non-sensitive data and
its revealed version; appropriate measures depend on the data type,
statistics, and the application as illustrated in the sequel.\ 

\emph{At its crux, our proposed sanitization process is about determining the
statistics of the output (database) that achieve a desired level of utility
and privacy and about deciding which input values to perturb and how to
probabilistically perturb them. Since the output statistics depends on the
sanitization process, for the i.i.d. source model considered here,
mathematically the problem reduces to finding the input to output symbol-wise
transition probability.}

\subsection{Background:\ Rate-distortion\ Theory}

In addition to a statistical model for large data sets, we also introduce an
abstract formulation for the sanitization process, which is based on the
theory of rate-distortion. We provide some intuition for the two steps
involved in information-theoretic sanitization, namely encoding at the
database and decoding at the data user.

For the purposes of privacy modeling the attributes about any individual in a
database fall in two categories: public attributes that can be revealed and
private attributes that need to be kept hidden, respectively. An attribute can
be both public and private at the same time. The attributes of any individual
are correlated; this implies that if the public attributes are revealed as is,
information about the private attributes can be inferred by the user using a
correlation model. Thus, ensuring privacy of the private attributes (also
referred to as hidden attributes in the sequel) requires
modifying/sanitizing/distorting the public attributes. However, the public
attributes have a utility constraint that limits the distortion, and
therefore, the privacy that can be guaranteed to the private attributes.

\begin{figure}[ptb]%
\centering
\includegraphics[
height=2.0159in,
width=2.3367in
]%
{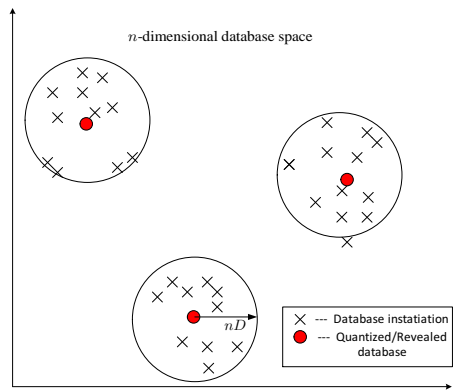}%
\caption{Space of all database realizations and the quantized databases. }%
\label{Fig_VQ}%
\end{figure}

Our approach is to determine the optimal sanitization, i.e., a mapping which
guarantees the maximal privacy for the private attributes for the desired
level of utility for the public attributes, among the set of \textit{all}
possible mappings that transform the public attributes of a database. We use
the terms \textit{encoding} and \textit{decoding} to denote this mapping at
the data publisher end and the user end respectively. A database instance is
an $n$-realization of a random source (the source is a vector when the number
of attributes $K>1)$ and can be viewed as a point in an $n$-dimensional space
(see Fig. \ref{Fig_VQ}). The set of all possible databases ($n$-length source
sequences) that can be generated using the source statistics (probability
distribution) lie in this space.

Our choice of utility metric is a measure of average `closeness' between the
original and revealed database public attributes via a distortion requirement
$D$. Thus the output of sanitization will be another database (another point
in the same $n$-dimensional space) within a ball of `distance' $nD$. We seek
to determine a set of some $M=2^{nR}$ output databases that `cover' the space,
i.e., given any input database instance there exists at least one sanitized
database within bounded `distance' $nD$ as shown in Fig. \ref{Fig_VQ}. Note
that the sanitized database may be in a subspace of the entire space because
only the public attributes are sanitized and the utility requirement is only
in this subspace.

In information theory such a distortion-constrained encoding is referred to as
quantization or compression. Furthermore, the mapping is referred to as vector
quantization because the compression is of an $n$-dimensional space and can be
achieved in practice using clustering algorithms. In addition to a distortion
(utility) constraint, our privacy constraint also requires that the
\textquotedblleft leakage\textquotedblright\ (i.e.\ the loss of uncertainty)
about the private attributes via correlation from the sanitized database is
bounded. The set of $M$ source-sanitized database pairs is chosen to satisfy
both distortion and leakage constraints. The database user that receives the
sanitized database may have other side-information (s.i.) about which the
encoder is either \textit{statistically} \textit{informed} (i.e., only the
statistics of s.i. known) or \textit{informed} (knows s.i. \textit{a priori}).
The decoder can combine the sanitized database published by the encoder and
the s.i. to recreate the final reconstructed database.

Obtaining the U-P tradeoff region involves two parts: the first is a proof of
existence of a mapping, called a \textit{converse} or outer bounds in
information theory, and the second is an \textit{achievable scheme} (inner
bounds) that involves constructing a mapping (called a code). Mathematically,
the converse bounds the maximal privacy that can be achieved for a desired
utility over the space of all feasible mappings, and the achievable scheme
determines the input to output probabilistic mapping and reveals the minimal
privacy achievable for a desired distortion. When the inner and outer bounds
meet, the constructive scheme is tight and achieves the entire U-P tradeoff,
often the case for tractable distributions such as Gaussian, Laplacian, and
arbitrary discrete sources.

It is important to note that our assumption of knowledge of the source
statistics at all involved parties does not limit the applicability of the
framework for the following reasons: (a) the statistics for large data can
often be sampled reliably from the data collected; (ii) knowledge of
statistics alone is insufficient to generate the actual database at the user;
and (iii) most importantly, the statistical knowledge enables us to find the
optimal input to output probabilistic mapping (i.e., a perturbation matched to
the source statistics) that satisfy specific utility and privacy measures. The
power of our approach is that it completely eliminates signal-perturbation
mismatch problems as observed in privacy-preserving data mining solutions by
Kargupta et al \cite{Kargupta}; furthermore, the irreversibility of the
quantization process implies that the suppressed or distorted data cannot be
reversed despite knowledge of the actual statistics. In the following Section,
we formalize these notions and present a rigorous analysis.%

\section{\label{Sec_Mod}Model and Metrics}

\subsection{Model for Databases}

A database $\mathcal{D}$ is a matrix whose rows and columns represent the
individual entries and their attributes, respectively. For example, the
attributes of a healthcare database can include name, address, SSN, gender,
and a collection of possible medical information. The attributes that directly
give away information such as name and SSN are typically considered private data.

\textit{Model}: Our proposed model focuses on large databases with $K$
attributes per entry. Let $\mathcal{X}_{k}$, for all $k$ $\in$ $\mathcal{K}%
=\left\{  1,2,,\ldots,K\right\}  $, and $\mathcal{Z}$ be finite sets. Let
$X_{k}\in\mathcal{X}_{k}$ be a random variable denoting the $k^{th}$
attribute, $k=1,2,\ldots,K,$ and let $X_{\mathcal{K}}\equiv\left(  X_{1}%
,X_{2},\ldots,X_{K}\right)  $. A database $d$ with $n$ rows is a sequence of
$n$ independent observations from the distribution having a probability
distribution%
\begin{equation}
p_{X_{\mathcal{K}}}\left(  x_{\mathcal{K}}\right)  =p_{X_{1}X_{2}\ldots X_{K}%
}\left(  x_{1},x_{2},\ldots,x_{K}\right)  \label{Prob_JointDist}%
\end{equation}
which is assumed to be known to both the designers and users of the database.
Our simplifying assumption of row independence holds generally in large
databases (but not always) as correlation typically arises across attributes
and can be ignored across entries given the size of the database. We write
$X_{\mathcal{K}}^{n}=\left(  X_{1}^{n},X_{2}^{n},\ldots,X_{K}^{n}\right)  $ to
denote the $n$ independent and identically distributed (i.i.d.) observations
of $X_{\mathcal{K}}^{n}$.

The joint distribution in (\ref{Prob_JointDist}) models the fact that the
attributes corresponding to an individual entry are correlated in general and
consequently can reveal information about one another.

\textit{Public and private attributes}: We consider a general model in which
some attributes need to be kept private while the source can reveal a function
of some or all of the attributes. We write $\mathcal{K}_{r}$ and
$\mathcal{K}_{h}$ to denote sets of private (subscript $h$ for hidden) and
public (subscript $r$ for revealed) attributes, respectively, such that
$\mathcal{K}_{r}\cup\mathcal{K}_{h}=\mathcal{K\equiv}\left\{  1,2,\ldots
,K\right\}  $. We further denote the corresponding collections of public and
private attributes by $X_{\mathcal{K}_{r}}\equiv\left\{  X_{k}\right\}
_{k\in\mathcal{K}_{r}}$ and $X_{\mathcal{K}_{h}}\equiv\left\{  X_{k}\right\}
_{k\in\mathcal{K}_{h}}$, respectively. More generally, we write
$X_{\mathcal{S}_{h}}\equiv\left\{  X_{k}:k\in\mathcal{S}_{h}\subseteq
\mathcal{K}_{h}\right\}  $ and $X_{\mathcal{S}_{r}}\equiv\left\{  X_{k}%
:k\in\mathcal{S}_{r}\subseteq\mathcal{K}_{r}\right\}  $ to denote subsets of
private and public attributes, respectively.

Our notation allows for an attribute to be both public and private; this is to
account for the fact that a database may need to reveal a function of an
attribute while keeping the attribute itself private. In general, a database
can choose to keep public (or private) one or more attributes ($K>1)$.
Irrespective of the number of private attributes, a non-zero utility results
only when the database reveals an appropriate function of some or all of its attributes.

\textit{Revealed attributes and side information}: As discussed in the
previous section, the public attributes are in general sanitized/distorted
prior to being revealed in order to reduce possible inferences about the
private attributes. We denote the resulting \textit{revealed attributes} as
$\hat{X}_{\mathcal{K}_{r}}\equiv\{\hat{X}_{k}\}_{k\in\mathcal{K}_{r}}$. In
addition to the revealed information, a user of a database can have access to
correlated side information from other information sources. We model the side
information (s.i.) as an $n$-length sequence $Z^{n}=\left(  Z_{1},Z_{2}%
,\ldots,Z_{n}\right)  $, $Z_{i}\in\mathcal{Z}$ for all $i,$ which is
correlated with the database entries via a joint distribution
$p_{X_{\mathcal{K}}Z}\left(  x_{\mathcal{K}}\mathbf{,}z\right)  $.

\textit{Reconstructed database}: The final \textit{reconstructed database} at
the user will be either a database of revealed public attributes (when no s.i.
is available) or a database generated from a combination of the revealed
public attributes and the side information (when s.i. is available).

\subsection{Metrics: The Privacy and Utility Principle}

Even though utility and privacy measures tend to be specific to the
application, there is a fundamental principle that unifies all these measures
in the abstract domain. A user perceives the utility of a perturbed database
to be high as long as the response is similar to the response of the
unperturbed database; thus, the utility is highest of an unperturbed database
and goes to zero when the perturbed database is completely unrelated to the
original database. Accordingly, our utility metric is an appropriately chosen
average `distance' function between the original and the perturbed databases.

Privacy, on the other hand, is maximized when the perturbed response is
completely independent of the data. Our privacy metric measures the difficulty
of extracting any private information from the response, i.e., the amount of
uncertainty or \textit{equivocation }about the private attributes given the
response. One could alternately quantify the \textit{privacy loss} from
revealing data as the \textit{mutual information} between the private
attributes and the response; mutual information is typically used to quantify
leakage (or secrecy) for continuous valued data.

\subsection{Utility and Privacy Aware Encoding}

Since database sanitization is traditionally the process of distorting the
data to achieve some measure of privacy, it is a problem of mapping a database
to a different one subject to specific utility and privacy requirements.

\textit{Mapping}: Our notation below relies on this abstraction.\ Let
$\mathcal{X}_{k},k\in\mathcal{K},$ and $\mathcal{Z},$ be as above and let
$\mathcal{\hat{X}}_{j}$ be additional finite sets for all $j\in\mathcal{K}%
_{r}$. Recall that a database $d$ with $n$ rows is an instantiation of
$X_{\mathcal{K}}^{n}$. Thus, we will henceforth refer to a real database $d$
as an \textit{input database} and to the corresponding sanitized database
(SDB) $d_{s}$ as an \textit{output database}. When the user has access to side
information, the \textit{reconstructed database} $d^{\prime}$ at the user will
in general be different from the output database.

Our coding scheme consists of an encoder $F_{E}$ which is a mapping from the
set of all input databases (i.e., all databases $d$ allowable by the
underlying distribution$)$ to a set of indices~$\mathcal{J}\equiv\left\{
1,2,\ldots,M\right\}  $ and an associated table of output databases (each of
which is a $d_{s})$ given by%
\begin{equation}
F_{E}:\left(  \mathcal{X}_{1}^{n}\times\ldots\times\mathcal{X}_{k}^{n}\right)
_{k\in\mathcal{K}_{enc}}\rightarrow\mathcal{J}\equiv\left\{  SDB_{k}\right\}
_{k=1}^{M} \label{F_Enc}%
\end{equation}
where $\mathcal{K}_{r}\subseteq\mathcal{K}_{enc}\subseteq\mathcal{K}$ and $M$
is the number of output (sanitized) databases created from the set of all
input databases. To allow for the case where an attribute can be both public
and private, we allow the encoding $F_{E}$ in (\ref{F_Enc}) to include both
public and private attributes. A user with a view of the SDB (i.e., an index
$j\in\mathcal{J})$ and with access to side information $Z^{n}$, whose entries
$Z_{i}$, $i=1,2,\ldots,n,$ take values in the alphabet $\mathcal{Z}$,
reconstructs the database $d^{\prime}$ via the mapping%
\begin{equation}
F_{D}:\mathcal{J}\times\mathcal{Z}^{n}\rightarrow\left(
{\textstyle\prod\nolimits_{k\in\mathcal{K}_{r}}}
\mathcal{\hat{X}}_{k}^{n}\right)  . \label{F_Dec}%
\end{equation}
The encoding and decoding are assumed known at both parties.

\textit{Utility}: Relying on a distance based utility principle, we model the
utility $u$ via the requirement that the average \textit{distortion} of the
public variables is upper bounded, for each $\epsilon>0$ and all sufficiently
large $n$, as
\begin{equation}
u\equiv\mathbb{E}\left[  \frac{1}{n}%
{\textstyle\sum_{i=1}^{n}}
\rho\left(  X_{\mathcal{K}_{r},i},\hat{X}_{\mathcal{K}_{r},i}\right)  \right]
\leq D+\epsilon\text{,} \label{Utility_mod}%
\end{equation}
where $\rho\left(  \cdot,\cdot\right)  $ denotes a distortion function,
$\mathbb{E}$ is the expectation over the joint distribution of
$(X_{\mathcal{K}_{r}},\hat{X}_{\mathcal{K}_{r}})$, and the subscript $i$
denotes the $i^{th}$ entry of the database. Examples of distortion functions
include the Euclidean distance for Gaussian distributions, the Hamming
distance for binary input and output databases, and the Kullback-Leibler (K-L)
divergence. We assume that $D$ takes values in a closed compact set to ensure
that the maximal and minimal distortions are finite and all possible
distortion values between these extremes can be achieved.

\textit{Privacy}: We quantify the equivocation $e$ of all the private
variables using entropy as%
\begin{equation}
e\equiv\frac{1}{n}H\left(  X_{\mathcal{K}_{h}}^{n}|J,Z^{n}\right)  \geq
E-\epsilon. \label{equivoc}%
\end{equation}
Analogous to (\ref{equivoc}), we can quantify the privacy leakage $l$ using
mutual information as
\begin{equation}
l\equiv\frac{1}{n}I\left(  X_{\mathcal{K}_{h}}^{n};J,Z^{n}\right)  \leq
L+\epsilon. \label{Leakage}%
\end{equation}

\begin{remark}
The case in which side information is not available at the user is obtained by
simply setting $Z^{n}=\emptyset$ in (\ref{F_Dec}) and (\ref{equivoc}).
\end{remark}

We shall henceforth focus on using equivocation as a privacy metric except for
the case where the source is modeled as continuous valued data since unlike
differential entropy, mutual information is strictly non-negative. From
(\ref{equivoc}), we have $H(X_{\mathcal{K}_{h}}|X_{\mathcal{K}r},Z)\leq E\leq
H(X_{\mathcal{K}_{h}}|Z)\leq H(X_{\mathcal{K}_{h}})$, where the upper bound on
the equivocation results when the private and public attributes (and side
information) are uncorrelated and the lower bound results when the public
attributes (and side information) completely preserve the correlation between
the public and private attributes. Note that the leakage can be analogously
bound as $0\leq I(X_{\mathcal{K}_{h}};Z)\leq L\leq I(X_{\mathcal{K}_{h}%
};X_{\mathcal{K}r},Z).$

The mappings in (\ref{F_Enc}) and (\ref{F_Dec}) ensure that $d$ is mapped to
$d^{\prime}$ such that the U-P constraints in (\ref{Utility_mod}) and
(\ref{equivoc}) are met. The formalism in (\ref{Prob_JointDist}%
)-(\ref{Leakage}) is analogous to lossy compression in that a source database
is mapped to one of $M$ quantized databases that are designed \textit{a
priori}. For a chosen encoding, a database realization is mapped to the
appropriate quantized database, subject to (\ref{Utility_mod}) and
(\ref{equivoc}). It suffices to communicate the index $J$ of the resulting
quantized database as formalized in (\ref{F_Enc}) to the user. This index, in
conjunction with side information, if any, enables a reconstruction at the
user as in (\ref{F_Dec}). \emph{Note that the mappings in (\ref{F_Enc}) and
(\ref{F_Dec}), i.e., lossy compression with privacy guarantees, ensure that
for any }$D>0$\emph{, the user can only reconstruct the database }%
$d^{^{\prime}}=\hat{X}_{\mathcal{K}_{r}}^{n}$\emph{, formally a function
}$f\left(  J,Z^{n}\right)  $\emph{, and not }$d=X_{\mathcal{K}}^{n}$\emph{
itself.}

The utility and privacy metrics in (\ref{Utility_mod}) and (\ref{equivoc})
capture the statistical nature of the problem, i.e., the fact that the entries
of the database statistically mirror the distribution (\ref{Prob_JointDist}).
Thus, both metrics represent averages across all database instantiations $d$,
and hence, (assuming stationarity and large $n$) over the sample space of
$X_{\mathcal{K}}$ thereby quantifying the average distortion (utility) and
equivocation (privacy) achievable per entry.

\begin{remark}
In general, a database may need to satisfy utility constraints for any
collection of subsets $\mathcal{S}_{r}^{\left(  l\right)  }$ $\subseteq
\mathcal{K}_{r}$ of attributes and privacy constraints on all possible subsets
of private attributes $\mathcal{S}_{h}^{\left(  m\right)  },$ $m=1,2,\ldots
,L_{p},$ $1\leq L_{p}\leq2^{\left\vert \mathcal{K}_{h}\right\vert }-1$ where
$\left\vert \mathcal{K}_{h}\right\vert $ is the cardinality of $\mathcal{K}%
_{h}$. For ease of exposition and without loss of generality, we develop the
results for the case of utility and privacy constraints on the set of all
public and private attributes. The results can be generalized in a
straightforward manner to constraints on arbitrary subsets.
\end{remark}

\section{\label{Sec_UPT}Utility-Privacy Tradeoffs}

Mapping utility to distortion and privacy to information uncertainty via
entropy (or leakage via mutual information) leads to the following definition
of the U-P tradeoff region.

\begin{definition}
\label{Def_UPT}The U-P tradeoff region $\mathcal{T}$ is the set of all
feasible U-P tuples $(D,E)$ for which there exists a coding scheme $\left(
F_{E},F_{D}\right)  $ given by (\ref{F_Enc}) and (\ref{F_Dec}), respectively,
with parameters $(n,M,u,e)$ satisfying the constraints in (\ref{Utility_mod})
and (\ref{equivoc}).
\end{definition}

While the U-P tradeoff region in Definition \ref{Def_UPT} can be determined
for specific database examples, one has to, in general, resort to numerical
techniques to solve the optimization problem \cite{DBPriv2}. To obtain closed
form solutions that define the set of all tradeoff points and identify the
optimal encoding schemes, we exploit the rich set of techniques from rate
distortion theory with and without equivocation constraints. To this end, we
study a more general problem of RDE by introducing an additional rate
constraint $M\leq2^{n\left(  R+\epsilon\right)  }$ which bounds the number of
quantized SDBs in (\ref{F_Enc}). Besides enabling the use of known
rate-distortion techniques, the rate constraint also has an operational
significance. For a desired level of accuracy (utility) $D$, the rate $R$ is
the precision required on average (over $\mathcal{X}_{K})$ to achieve it. We
now define the achievable RDE region as follows.

\begin{definition}
\label{Def_RDE}The RDE region $\mathcal{R}_{RDE}$ is the set of all tuples
$(R,D,E)$ for which there exists a coding scheme given by (\ref{F_Enc}) and
(\ref{F_Dec}) with parameters $(n,M,u,e)$ satisfying the constraints in
(\ref{Utility_mod}), (\ref{equivoc}), and on the rate. In this region,
$\mathcal{R}_{D-E}$, the set of all feasible distortion-equivocation tuples
$\left(  D,E\right)  $ is defined as
\begin{equation}
\mathcal{R}_{D-E}\equiv\left\{  (D,E):\left(  R,D,E\right)  \in\mathcal{R}%
_{RDE}\text{, }R\geq0\right\}  .
\end{equation}

\end{definition}

The RDE problem differs from the distortion-equivocation problem in including
a constraint on the precision of the public variables in addition to the
equivocation constraint on the private data in both problems. Thus, in the RDE
problem, for a desired utility $D$, one obtains the set of all
rate-equivocation tradeoff points $\left(  R,E\right)  ,$ and therefore, over
all distortion choices, the resulting region contains the set of all $\left(
D,E\right)  $ pairs. From Definitions \ref{Def_UPT} and \ref{Def_RDE}, we thus
have the following proposition.

\begin{proposition}
\label{Prop1}$\mathcal{T}=\mathcal{R}_{D-E}$.
\end{proposition}

\begin{figure*}[tbp] \centering
{\includegraphics[
height=2.7363in,
width=5.5002in
]%
{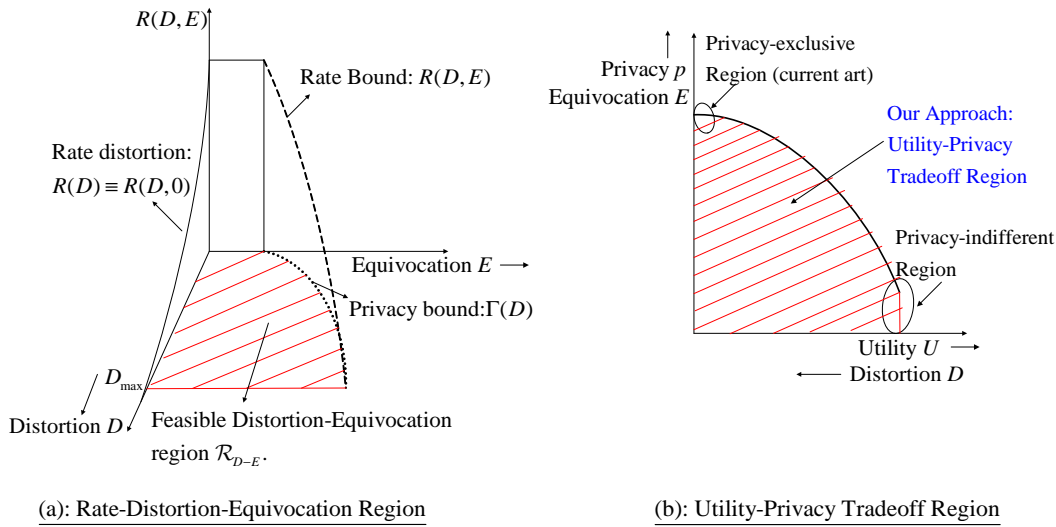}%
}
\caption{(a) Rate Distortion Equivocation Region \cite{Yamamoto}; (b) Utility-Privacy Tradeoff Region.}\label{Fig_RDE_UP}%
\end{figure*}

Proposition \ref{Prop1} is captured pictorially in Fig. \ref{Fig_RDE_UP}(b).
The functions $R\left(  D,E\right)  $ and $\Gamma(D)$ in Fig. \ref{Fig_RDE_UP}
capture the rate and privacy boundaries of the region and are the minimal rate
and maximal privacy achievable, respectively, for a given distortion $D$.%

The power of Proposition \ref{Prop1} is that it allows us to study the larger
problem of database U-P tradeoffs in terms of a relatively familiar problem of
source coding with additional privacy constraints. Our result shows the
tradeoff between utility (distortion), privacy (equivocation), and precision
(rate) --\ fixing the value of any one determines the set of operating points
for the other two; for example, fixing the utility (distortion $D$) quantifies
the set of all achievable privacy-precision tuples $\left(  E,R\right)  $.

For the case of no side information, i.e., for the problem in (\ref{F_Enc}%
)-(\ref{equivoc}) with $Z^{n}=\emptyset$, the RDE region was obtained by
Yamamoto \cite{Yamamoto} for $K_{r}=K_{h}=1$ and $\mathcal{K}_{r}%
\cap\mathcal{K}_{h}=\emptyset$. We henceforth refer to this as an
\textit{uninformed case}, since neither the encoder (database) nor the decoder
(user) have access to external information sources. We summarize the result
below in the context of a utility-privacy tradeoff region. We first summarize
the intuition behind the results and the encoding scheme achieving it.

In general, to obtain the set of all achievable RDE tuples, one follows two
steps: the first is to obtain (\textit{outer) bounds }for a $\left(
n,M,u,e\right)  $ code on the rate and equivocation required to decode
reliably with a distortion $D$ (vanishing error probability in decoding for a
bounded distortion $D$); the second step is a constructive coding scheme for
which one determines the \textit{inner bounds} on rate and equivocation. The
set of all $\left(  R,D,E\right)  $ tuples is achievable when the two bounds
meet. The achievable RDE region was developed in \cite[Appendix]{Yamamoto} for
the problem in \ref{Def_RDE}. Focusing on the set of all RDE tradeoff points,
we restate the results in \cite[Appendix]{Yamamoto} as follows.

\begin{proposition}
\label{Prop_UI}Given a database with public, private, and reconstructed
variables $X_{\mathcal{K}_{r}},$ $X_{\mathcal{K}_{h}}$, and $\hat
{X}_{\mathcal{K}_{r}}$ respectively, and $Z=\emptyset,$ for a fixed target
distortion $D$, the set of achievable $(R,E)$ tuples satisfy
\begin{subequations}
\label{RDE_Yam}%
\begin{align}
R  &  \geq R_{U}\left(  D\right)  \equiv I(X_{\mathcal{K}_{r}}X_{\mathcal{K}%
_{h}};\hat{X}_{\mathcal{K}_{r}})\label{R_U_D}\\
E  &  \leq E_{U}\left(  D\right)  \equiv H(X_{\mathcal{K}_{h}}|\hat
{X}_{\mathcal{K}_{r}}) \label{E_U_D}%
\end{align}
for some $p(x_{\mathcal{K}_{h}},x_{\mathcal{K}_{r}},\hat{x}_{\mathcal{K}_{r}%
})$ such that $\mathbb{E}(d(X_{\mathcal{K}_{r}},\hat{X}_{\mathcal{K}_{r}%
}))\leq D$.
\end{subequations}
\end{proposition}

\begin{remark}
The distribution $p(x_{\mathcal{K}_{h}},x_{\mathcal{K}_{r}},\hat
{x}_{\mathcal{K}_{r}})$ allows for two cases, one in which both the public and
private attributes are used to encode (e.g., medical) and the other in which
only the public (e.g., census) attributes are used. For the latter case in
which the private attributes are only implicitly used (via the correlation),
the distribution simplifies as $p(x_{\mathcal{K}_{h}},x_{\mathcal{K}_{r}%
})p(\hat{x}_{\mathcal{K}_{r}}|x_{\mathcal{K}_{h}})$, i.e., the variables
satisfy the Markov chain $X_{\mathcal{K}_{h}}-X_{\mathcal{K}_{r}}-\hat
{X}_{\mathcal{K}_{r}}$.
\end{remark}

\begin{theorem}
\label{TheoremUninformed}The U-P tradeoff region for a database problem
defined by (\ref{Prob_JointDist})-(\ref{equivoc}) and with $Z^{n}=\emptyset$
is the set of all $\left(  E,D\right)  $ such that for every choice of
distortion $D\in\mathcal{D}$ that is achievable by quantization scheme with a
distribution $p(x_{\mathcal{K}_{h}},x_{\mathcal{K}_{r}}\hat{x}_{\mathcal{K}%
_{r}}),$ the privacy achievable is given by $E_{U}(D)$ in (\ref{E_U_D}) (for
which a rate of $R_{U}\left(  D\right)  $ in (\ref{R_U_D}) is required).
\end{theorem}

The set of all RDE tuples in (\ref{RDE_Yam}) define the region $\mathcal{R}%
_{RDE}^{\ast}.$ The functions in Fig. \ref{Fig_RDE_UP} specifying the
boundaries of this region are given as follows: $R\left(  D,E\right)  $ which
is the minimal rate required for any choice of distortion $D$ is given by
\begin{equation}
R\left(  D,E\right)  =R\left(  D,E^{\ast}\right)  =\min_{p(x_{\mathcal{K}_{h}%
},x_{\mathcal{K}_{r}},\hat{x}_{\mathcal{K}_{r}})}R_{U}\left(  D\right)  ,
\label{Rmin_Yam}%
\end{equation}
where $E^{\ast}=E_{U}(D)|_{p^{\ast}}$ is evaluated at $p^{\ast}$ is the
argument of the optimization in (\ref{Rmin_Yam}) and $\Gamma(D)$ which is the
maximal equivocation achievable for a desired distortion $D$ is given by%
\begin{equation}
\Gamma(D)=\max_{\min_{p(x_{\mathcal{K}_{h}},x_{\mathcal{K}_{r}},\hat
{x}_{\mathcal{K}_{r}})}}E_{U}\left(  D\right)  .
\end{equation}

\begin{remark}
In general, the functions $R\left(  D,E\right)  $ and $\Gamma\left(  D\right)
$ may not be optimized by the same distribution $p(x_{\mathcal{K}_{h}%
},x_{\mathcal{K}_{r}},\hat{x}_{\mathcal{K}_{r}})$, i.e., $R\left(  D,E\right)
$ may be minimal for a $E=E^{\ast}<\Gamma(D)$. This implies that in general
the minimal rate encoding scheme is not necessarily the same as the encoding
scheme that maximizes equivocation (privacy) for a given distortion $D$. This
is because a compression scheme that only satisfies a fidelity constraint on
$X_{\mathcal{K}_{r}}$, i.e., source coding without additional privacy
constraints, is oblivious of the resulting leakage of $X_{\mathcal{K}_{h}}$
whereas a compression scheme which minimizes the leakage of $X_{\mathcal{K}%
_{h}}$ while revealing $X_{\mathcal{K}_{r}}$ will first reveal that part of
$X_{\mathcal{K}_{r}}$ that is orthogonal to $X_{\mathcal{K}_{h}}$ and only
reveal $X_{\mathcal{K}_{h}}$ when the fidelity requirements are high enough to
encode it. Thus, maximal privacy may require additional precision (of the
component of $X_{\mathcal{K}_{r}}$ orthogonal to $X_{\mathcal{K}_{h}})$
relative to the fidelity-only case. The additional rate constraint enables us
to intuitively understand the nature of the lossy compression scheme required
when privacy need to be guaranteed.
\end{remark}

We now focus on the case in which the user has access to correlated side
information. The resulting RDE tradeoff theorems generalize the results in
\cite{Yamamoto}; furthermore, we present a new relatively easier proof for the
achievable equivocation while introducing a class of encoding schemes that we
refer to as \textit{quantize-and-bin coding }(see also \cite{LS_DBPriv3}).

\subsection{\label{Sec_SI}Capturing the Effects of Side-Information}

In general, a user can have access to auxiliary information either from prior
interactions with the database or from a correlated external source. We cast
this problem in information-theoretic terms as a database encoding problem
with side information at the user. Two cases arise in this context: i) the
database has knowledge of the side information due to prior interactions with
the user and is sharing a related but differently sanitized view in the
current interaction, i.e., an \textit{informed encoder}; and ii) the database
does not know the exact side information but has some statistical knowledge,
i.e., an \textit{statistically informed encoder}. \ We develop the RDE regions
for both cases below.

\subsubsection{U-P\ Tradeoffs: Statistically Informed Encoder}

We first focus on the case with side information at the user and knowledge of
its statistics at the encoder, i.e., at the database. The following theorem
quantifies the RDE\ region, and hence, the utility-privacy tradeoff region for
this case.

\begin{theorem}
\label{TheoremS_informed} For a target distortion $D$, the set of achievable
$(R,E)$ tuples when the database has access to the statistics of the side
information is given as
\begin{subequations}
\label{RDE_WZ}%
\begin{align}
R  &  \geq R_{SI}\left(  D\right)  \equiv I(X_{\mathcal{K}_{r}}X_{\mathcal{K}%
_{h}};U|Z)\label{R_WZ}\\
E  &  \leq E_{SI}\left(  D\right)  \equiv H(X_{\mathcal{K}_{h}}|UZ)
\label{E_WZ}%
\end{align}
for some distribution $p(x_{\mathcal{K}_{h}},x_{\mathcal{K}_{r}},z)p\left(
u|x_{\mathcal{K}_{h}},x_{\mathcal{K}_{r}}\right)  $ such that there exists a
function $\hat{X}_{\mathcal{K}_{r}}=f(U,Z)$ for which $\mathbb{E}\left[
d(X_{\mathcal{K}_{r}},\hat{X}_{\mathcal{K}_{r}})\right]  \leq D$, and
$\left\vert \mathcal{U}\right\vert =\left\vert \mathcal{X}_{\mathcal{K}%
}\right\vert +1$.
\end{subequations}
\end{theorem}

\begin{remark}
For the case in which only the public variables are used in encoding, i.e.,
$X_{\mathcal{K}_{h}}-X_{\mathcal{K}_{r}}-U,$ $\left\vert \mathcal{U}%
\right\vert =\left\vert \mathcal{X}_{\mathcal{K}_{r}}\right\vert +1.$
\end{remark}

We prove Theorem \ref{TheoremS_informed} in the Appendix. Here, we present a
sketch of the achievability proof. The main idea is to show that a
quantize-and-bin encoding scheme achieves the RDE tradeoff.

The intuition behind the quantize-and-bin coding scheme is as follows: the
source $\left(  X_{\mathcal{K}_{r}}^{n},X_{\mathcal{K}_{h}}^{n}\right)  $ is
first quantized to $U^{n}$ at a rate of $I(X_{\mathcal{\dot{K}}_{r}%
}X_{\mathcal{K}_{h}}^{n};U)$. For the uninformed case, the encoder would have
simply sent the index for $U^{n}$ ($\equiv\hat{X}_{\mathcal{K}_{r}}^{n})$ to
the decoder. However, since the encoder has statistical knowledge of the
decoder's side information, the encoder further bins $U^{n}$ to reduce the
transmission rate to $I(X_{\mathcal{K}_{r}}X_{\mathcal{K}_{h}};U)-I(Z;U)$
where $I(Z;U)$ is a measure of the correlation between $Z^{n}$ and $U^{n}$.
The encoder then transmits this bin index $J$ so that using $J$ and $Z^{n}$,
the user can losslessly reconstruct $U^{n},$ and hence, $\hat{X}%
_{\mathcal{K}_{r}}^{n}=f\left(  U^{n},Z^{n}\right)  $ via a deterministic
function $f$ to the desired $D$.

The outer bounds follow along the lines of the Wyner-Ziv converse as well as
outer bounds on the equivocation (see the\ Appendix). The key result here is
the inner bound on the equivocation, i.e., for a fixed distortion $D,$ the
quantize-and-bin encoding scheme can guarantee a lower bound on the
equivocation as $H(X_{\mathcal{K}_{h}}|U,Z)$ which primarily relies on the
fact that using the bin index $J$ and side information $Z^{n}$, the quantized
database $U^{n}$ can be losslessly reconstructed at the user.

\textit{Uninformed case}: Here, we have $Z=0$ and $U=\hat{X}_{\mathcal{K}_{r}%
}$, i.e., the reconstructed and sanitized databases are the same. Note that in
this case, the quantize-and-bin scheme simplifies to a simple quantize scheme
(as required to achieve Proposition \ref{Prop_UI}).

\begin{remark}
For a desired $D,$ minimizing $R_{SI}(D)$ yields the Wyner-Ziv rate-distortion
function. However, we focus here on the tradeoff region, and hence, the set of
\textit{all} $\left(  R,D,E\right)  $ tuples.
\end{remark}

\subsubsection{U-P\ Tradeoffs: Informed Encoder}

We now consider the case in which the encoder also has perfect knowledge of
the side information. Such a case can arise in practice if the encoder has
shared some prior information related to the database earlier. The following
theorem summarizes the RDE tradeoff region for this case.

\begin{theorem}
\label{Theorem_Inf} For a target distortion $D$, the set of achievable $(R,E)$
tuples when the encoder has perfect knowledge of the side information is given
as
\begin{subequations}
\label{RDE_HB}%
\begin{align}
R  &  \geq R_{I}\left(  D\right)  \equiv I(X_{\mathcal{K}_{r}},X_{\mathcal{K}%
_{h}};\hat{X}_{\mathcal{K}_{r}}|Z)\\
E  &  \leq E_{I}\left(  D\right)  \equiv H(X_{\mathcal{K}_{h}}|\hat
{X}_{\mathcal{K}_{r}}Z)
\end{align}
for some distribution $p(x_{\mathcal{K}_{h}},x_{\mathcal{K}_{r}},z)p\left(
\hat{x}_{\mathcal{K}_{r}}|x_{\mathcal{K}_{h}},x_{\mathcal{K}_{r}},z\right)  $
for which $\mathbb{E}\left[  d(X_{\mathcal{K}_{r}},\hat{X}_{\mathcal{K}_{r}%
})\right]  \leq D$.
\end{subequations}
\end{theorem}

\begin{remark}
For $Z^{n}=\emptyset,$ Theorem \ref{Theorem_Inf} simplifies to Proposition
\ref{Prop_UI}.
\end{remark}

We prove Theorem \ref{Theorem_Inf} in the Appendix. The main idea is to show
that an \textit{informed quantize-and-bin} encoding scheme for the informed
case in which \emph{both} $(X_{\mathcal{K}}^{n},Z^{n})$ are available at the
encoder achieves the RDE tradeoff. The encoder jointly compresses them to a
database $\hat{X}_{\mathcal{K}_{r}}^{n}$ which it further bins and reveals the
bin index to the decoder such that the rate of transmission reduces to
$I(X_{\mathcal{K}}Z;\hat{X}_{\mathcal{K}_{r}})-I(Z;\hat{X}_{\mathcal{K}_{r}%
})=I(X_{\mathcal{K}};\hat{X}_{\mathcal{K}_{r}}|Z)$. Using the bin index and
side information $Z^{n}$, the database $\hat{X}_{\mathcal{K}_{r}}^{n}$ can be
losslessly reconstructed. The outer bounds follow from standard results on
conditional rate-distortion converse (see the\ Appendix). The key result is
the inner bound on the equivocation, i.e., for a fixed $D,$ the
quantize-and-forward scheme is shown to guarantee a minimal equivocation of
$H(X_{\mathcal{K}_{h}}|\hat{X}_{\mathcal{K}_{r}},Z)$ using the fact that from
$J$ and $Z^{n}$, $\hat{X}_{\mathcal{K}_{r}}^{n}$ can be losslessly
reconstructed at the user.

\section{\label{Sec_Illus}Illustration of Results}

In this Section, we apply the utility-privacy framework we have introduced to
model two fundamental types of databases and illustrate the corresponding
optimal coding schemes that achieve the set of all utility-privacy tradeoff
points. More importantly, we demonstrate how the optimal input to output
probabilistic mapping (coding scheme) in each case sheds light on practical
privacy-preserving techniques. We note that for the i.i.d. source model
considered, vector quantization (to determine the set of $M$ output databases)
simplifies to finding the probabilities of mapping the letters of the source
to letters of the output (database) alphabet as formally shown in the previous Section.

We model two broad classes of databases: \textit{categorical} and
\textit{numerical}. Categorical data are typically discrete data sets
comprising information such as gender, social security numbers and zip codes
that provide (meaningful) utility only if they are mapped within their own
set. On the other hand, without loss of generality, numerical data can be
assumed to belong to the set of real numbers or integers as appropriate. In
general, a database will have a mixture of categorical and numerical
attributes, but for the purpose of illustration, we assume that the database
is of one type or the other, i.e., every attribute is of the same kind. In
both cases, we assume a single utility (distortion) function. We discuss each
example in detail below.

Recall that the abstract mapping in (\ref{F_Enc}) is a lossy compression of
the database. The underlying principle of optimal lossy compression is that
the number of bits required to represent a sample $x$ of $X\sim p_{X}$ is
inversely proportional to $\log\left(  p(x)\right)  $, and thus, for a desired
$D$, preserving the events in descending order of $p_{X}$ requires the least
number of bits on average. The intuitive notion of privacy as being
unidentifiable in a crowd is captured in this information-theoretic
formulation since the low probability entries, the \textit{outliers, }that
convey the most information, are the least represented. It is this fundamental
notion that is captured in both examples.

\begin{example}
Consider a categorical database with $K\geq1$ attributes. In general, the
$k^{th}$ attribute $X_{k}$ takes values in a discrete set $\mathcal{X}_{k}$ of
cardinality $M_{k}$. For our example, \emph{we assume that all attributes need
to be revealed}, and therefore, it suffices to view each entry (a row of all
$K$ attributes) of the database as generated from a discrete scalar source $X$
of cardinality $M$, i.e., $X\sim p(x),$ $x\in\left\{  1,2,\ldots,M\right\}  $.
Taking into account the fact that sanitizing categorical data requires mapping
within the same set, for this arbitrary discrete source model, we assume that
the output sample space $\mathcal{\hat{X}=X}$. Since changing a sample of the
categorical data can significantly change the utility of the data, we account
for this via a utility function that penalizes such changes. We thus model the
utility function as a generalized Hamming distortion which captures this cost
model (averaged over all samples of $X)$ such that the average distortion $D$
is given by
\begin{equation}
D=\Pr\left\{  X\not =\hat{X}\right\}  . \label{CatDB_Dist}%
\end{equation}
Focusing on the problem of revealing the entire database $d=X^{n}$ (a
$n$-sequence realization of $X)$ as $\hat{X}^{n}$, we define the equivocation
as
\begin{equation}
\frac{1}{n}H(X^{n}|\hat{X}^{n})\geq E. \label{CatEq}%
\end{equation}
Thus, the utility-privacy problem is that of finding the set of all $\left(
D,E\right)  $ pairs such that for every choice of $p(\hat{x}|x)$ achieving a
desired $D,$ the equivocation is bounded as in (\ref{CatEq}). Applying
Proposition \ref{Prop_UI} (and also Theorem \ref{Theorem_Inf} with
$Z^{n}=\emptyset)$, we have that for a target distortion $D$, the set of
achievable $(R,E)$ tuples satisfy
\begin{subequations}
\label{CatRE}%
\begin{equation}
R\geq R_{U}\left(  D\right)  \equiv I(X;\hat{X});\text{ }E\leq E_{U}\left(
D\right)  \equiv H(X|\hat{X})
\end{equation}
for some distribution $p(x)p\left(  \hat{x}|x\right)  $ for which
$\mathbb{E}\left[  d(X,\hat{X})\right]  \leq D$. Note that the rate
$R_{U}\left(  D\right)  =H(X)-E_{U}(D)$, and thus, minimizing $R_{U}\left(
D\right)  $ for a desired $D$ maximizes $E_{U}\left(  D\right)  .$ Thus, while
(\ref{CatRE}) defines the set of all $\left(  R,D,E\right)  $ tuples, we focus
on the $\left(  D,E\right)  $ pairs for which maximal equivocation (privacy)
is achieved.

The problem of minimizing $R_{U}\left(  D\right)  $ for an arbitrary source
with a generalized Hamming distortion has been studied in \cite{Pinkston} who
showed that $R(D)$ is achieved by reverse waterfilling solution such that
\end{subequations}
\begin{equation}
p(\hat{x})=\frac{\left(  p(x)-\lambda\right)  ^{+}}{\sum_{x\in\mathcal{X}%
}\left(  p(x)-\lambda\right)  ^{+}} \label{DBCat_pxhat}%
\end{equation}
and the `test channel' (mapping from $\hat{X}$ to $X)$ is given by%
\begin{equation}
p(x|\hat{x})=\left\{
\begin{array}
[c]{ll}%
\overline{D}, & x=\hat{x}\\
\lambda, & x\not =\hat{x},x\in\mathcal{\hat{X}}_{\text{supp}}\\
p_{k}, & x=k\not \in \mathcal{\hat{X}}_{\text{supp}}%
\end{array}
\right.  \label{DBCat_pxhatx}%
\end{equation}
where $\overline{D}=1-D$, $\lambda$ is chosen such that $\sum_{\hat{x}}%
p(\hat{x})p(x|\hat{x})=p(x)$, $p_{k}=p\left(  x=k\right)  $, and
$\mathcal{\hat{X}}_{\text{supp}}=\left\{  x:p(x)-\lambda>0\right\}  .$ Let
$S=\left\vert \mathcal{\hat{X}}_{\text{supp}}\right\vert -1.$ The maximal
achievable equivocation, and hence, the largest utility-privacy tradeoff
region is%
\begin{equation}
\Gamma(D)=-\overline{D}\log\overline{D}-S\lambda\log\lambda-\sum
_{k\not \in \mathcal{\hat{X}}_{\text{supp}}}p_{k}\log p_{k}.
\end{equation}
The waterlevel $\lambda$ is the Lagrangian for the distortion constraint in
minimizing $R_{U}\left(  D\right)  $. The distribution of entries in
$d^{^{\prime}}$ in (\ref{DBCat_pxhat}) demonstrates that the source samples
with low probabilities relative to the water level are not preserved, leading
to a `flattening' of the output distribution. Thus, we see that the commonly
used heuristics of outlier suppression, aggregation, and imputation
\cite{Raghu,Dobra} on census and related databases can be formally shown to
minimize privacy leakage for the appropriate model. We illustrate our results
in\ Fig. \ref{FigPMF} for $p_{X}\left(  x\right)  =[0.25$ $0.25$ $0.15$ $0.1$
$0.04$ $0.005$ $0.003$ $0.002]$ in which the first subplot demonstrates
increased suppression of the outliers with increasing $D,$ and the second
shows the entire U-P region.%
\begin{figure*}[tbp] \centering
{\includegraphics[
trim=0.134180in 0.075669in 0.000000in 0.077203in,
height=2.725in,
width=5.7778in
]%
{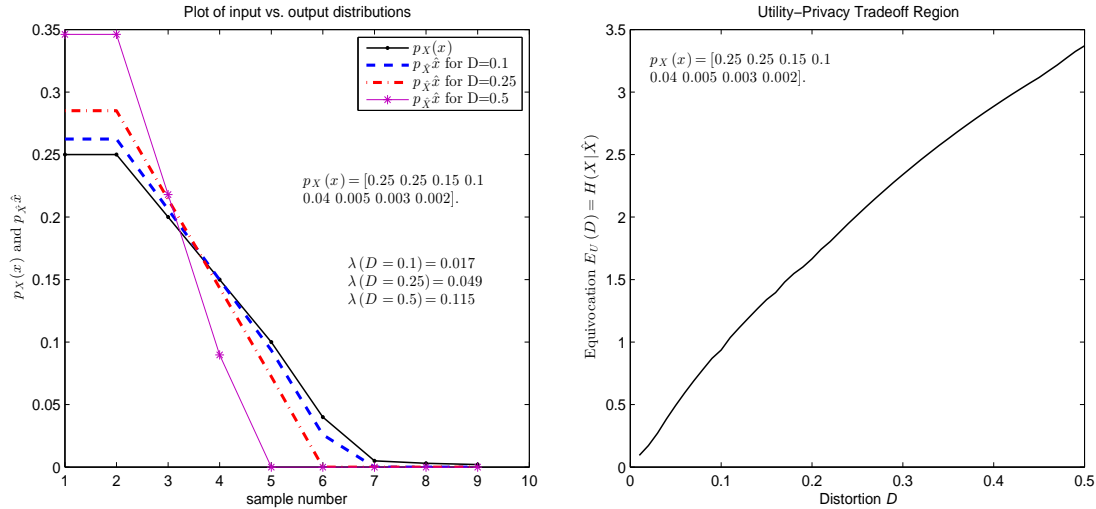}%
}
\caption{a) Reverse WF distributions for D=0.1,0.25,0.5; b) U-P tradeoff region.}\label{FigPMF}%
\end{figure*}%

\end{example}

\emph{Interpretation:} The probability $p(x)$ is the assumed probability of
occurrence of each unique sample (e.g., names such as Smith, Johnson, Poor,
Sankar, etc.) in the database. For categorical data, the attribute space for
the input and output databases are assumed to be the same (e.g., names mapped
to names). The Hamming distortion measure we have chosen quantifies the
average probability of a true sample of the source being mapped to a different
sample in the output database (e.g., probability that a name in the input
database is mapped to a different name in the output database averaged over
all names). The output distribution in (\ref{DBCat_pxhat}) implies that for a
desired utility (quantified via a Hamming distortion $D$), all the input
samples with probabilities below a certain $\lambda$ (e.g., say `Sankar,' a
very low probability name) will \emph{not} be present in the output database.
The water-level $\lambda$ is chosen such that the input and output database
samples satisfy $D$ in (\ref{CatDB_Dist}). Thus, the probability of guessing
that Sankar was in the original database given one only sees Smith, Johnson,
and Poor is given by (\ref{DBCat_pxhatx}) and is the same as the probability
of Sankar in the original database, i.e., there is no reduction in uncertainty
about Sankar given the published data! Furthermore, given that the name Smith
is published, the probability that Smith resulted from others such as Johnson,
Poor, and Sankar as well as from\ Smith is also given by (\ref{DBCat_pxhatx}).
This shows that every sample in the output database contains some uncertainty
about the actual sample with maximal uncertainty for those suppressed.
\textit{Our mapping not only mathematically minimizes the leakage of the
original samples but also does so to provide privacy to all and maximally to
those who are viewed as outliers (relative to the utility measure)}. For
simplicity, we have chosen a single private attribute, name, in this example.
In general, there could be several correlated attributes (e.g.\ name and last
four digits of the SSN) that will be changed together. This is captured by our
joint distribution. This eliminates the possibility that the adversary uses
his knowledge of the distribution to tell which individual entries have been
changed. The use of Hamming distortion measure in this example illustrates
another aspect of the power of our model. Sanitization of non-numeric data
attributes in a utility-preserving way is hard to do, especially because
distance metrics for non-numeric data tend to be application-specific. Hamming
distortion is an example of an extreme measure that penalizes every change
uniformly, no matter how small the change. It may be appropriate to use this
measure for applications that are especially sensitive to utility loss.

\begin{example}
In this example we model a numerical (e.g.\ medical) database in which the
attributes such as weight and blood pressure are often assumed to be normally
(Gaussian) distributed. Specifically, we consider a $K=2$ database with a
public $X$ ($\equiv X_{r})$ and a private $Y$ ($\equiv X_{h})$ attribute such
that $X$ and $Y$ are jointly Gaussian with zero means and variances
$\sigma_{X}^{2}$ and $\sigma_{Y}^{2}$, respectively, and a correlation
coefficient $\rho_{XY}=E\left[  XY\right]  /\left(  \sigma_{X}\sigma
_{Y}\right)  $. We assume that only $X$ is encoded such that $Y-X-\hat{X}$
holds. We consider three cases: (i) no side information, (ii) side information
$Z^{n}$ at user, and (iii) $Z^{n}$ at both. For the cases with $Z^{n}$, we
assume that $Z$ is i.i.d. zero mean with variance $\sigma_{Z}^{2}$ and is
jointly Gaussian with $(X,Y)$ such that $Y-X-Z$ forms a Markov chain and has a
correlation coefficient $\rho_{XZ}=E\left[  XZ\right]  /\left(  \sigma
_{X}\sigma_{Z}\right)  $. We use the leakage $L$ in (\ref{Leakage}) as the
privacy metric.

Case (i):\ No side information: The $(R,D,L)$ region for this case can be
obtained directly from Proposition \ref{Prop_UI} in (\ref{RDE_Yam}) with
$\hat{X}_{K_{r}}\equiv\hat{X}$ and $E_{U}\left(  D\right)  $ replaced by
$L_{U}\left(  D\right)  \equiv I(Y;\hat{X})$. For a Gaussian $\left(
X,Y\right)  ,$ one can easily verify that, for a desired $D,$ both $R_{U}(D)$
and $L_{U}(D)$ are minimized by a Gaussian $\hat{X}$ \cite[Chap. 10]{CTbook},
i.e., \textit{for normally distributed databases, the privacy-maximizing
revealed database is also normally distributed}. Furthermore, due to
$Y-X-\hat{X}$, the minimization of $I(X;\hat{X})$ is strictly over $p(\hat
{x}|x),$ and thus, simplifies to the familiar R-D problem for a Gaussian
source that is achieved by choosing $\hat{X}=X+N$, where the noise
$N\sim\mathcal{N}\left(  0,\sigma_{N}^{2}\right)  $ is independent of $X$ and
its variance $\sigma_{N}^{2}$ is chosen such that $D=Evar\left(  X|\hat
{X}\right)  \in\left[  0,\sigma_{X}^{2}\right]  $ where $var$ denotes
variance. The resulting minimal rate and leakage achieved (in bits per entry)
are, for $D\in\left[  0,\sigma_{X}^{2}\right]  ,$%
\begin{align*}
R_{U}^{\ast}\left(  D\right)   &  =\frac{1}{2}\log\left(  \frac{\sigma_{X}%
^{2}}{D}\right)  ,\\
L_{U}^{\ast}(D)  &  =\frac{1}{2}\log\left(  \frac{1}{\left[  \left(
1-\rho_{XY}^{2}\right)  +\rho_{XY}^{2}D\left/  \sigma_{X}^{2}\right.  \right]
}\right)  \text{.}%
\end{align*}
The largest U-P tradeoff region is thus the region enclosed by $L(D)$.

Case (ii): For the statistically informed encoder, the $\left(  R,D,L\right)
$ region is given by (\ref{RDE_WZ}) with $E_{SI}\left(  D\right)  $ replaced
by $L_{SI}\left(  D\right)  =I(Y;UZ).$ One can show the optimality of Gaussian
encoding in minimizing both the rate and leakage in \ref{RDE_WZ}, and thus, we
have $U=X+N,$ where $N\sim\mathcal{N}\left(  0,\sigma_{N}^{2}\right)  $ is
independent of $X$ and its variance $\sigma_{N}^{2}$ is chosen such that the
distortion $D=Evar\left(  X|UZ\right)  \in\left[  0,\sigma_{X}^{2}\right]  $.
Computing the minimal rate $R_{SI}^{\ast}\left(  D\right)  $ (the Wyner-Ziv
rate \cite{Wyner_Ziv}) and leakage $L_{SI}^{\ast}\left(  D\right)  $ for a
jointly Gaussian distribution achieving a distortion $D$, we obtain for all
$D\in\left[  0,\sigma_{X}^{2}\left(  1-\rho_{XZ}\right)  \right]  ,$%
\begin{align*}
R_{SI}^{\ast}\left(  D\right)   &  =R_{WZ}\left(  D\right)  =\frac{1}{2}%
\log\left(  \frac{\sigma_{X}^{2}\left(  1-\rho_{XZ}^{2}\right)  }{D}\right) \\
L_{SI}^{\ast}(D)  &  =L_{U}^{\ast}\left(  D\right)  ,
\end{align*}
i.e., the minimal rate and leakage are independent of $\rho_{XY}^{2}$ and
$\rho_{XZ}^{2}$, respectively, and thus, \textit{user side information does
not degrade privacy when the minimal-rate encoding is used}. The access to
side information at the user implies that the maximal achievable distortion is
at most as large as the uninformed case. Note that unlike $L_{U}^{\ast}\left(
D\right)  $ which goes to zero at the maximal distortion of $\sigma_{X}^{2}$,
$L_{SI}^{\ast}\left(  D\right)  >0$ for $D=\sigma_{X}^{2}\left(  1-\rho
_{XZ}^{2}\right)  $ as a result of the implicit correlation between $Y$ and
$Z$. These observations are clearly shown in Fig. \ref{Fig_Gauss} for
$\sigma_{X}^{2}=1$ and different values of $\rho_{XY}^{2}$ and $\rho_{XZ}^{2}$.

Case (iii): Finally, for a Gaussian source model, the $\left(  R,D,L\right)  $
region achievable for the informed encoder-decoder pair is the same as that
for Case (ii). This is because of the no rate-loss property of Wyner-Ziv
coding for a Gaussian source, i.e., knowledge of the side information
statistics at the encoder suffices to remove the correlation\ from each entry
before sharing data with the user \cite{Wyner_02}. Furthermore, since Gaussian
outputs minimize the rate as well as the leakage, the minimal $R_{I}^{\ast
}\left(  D\right)  =R_{SI}^{\ast}\left(  D\right)  $ and $L_{I}^{\ast}\left(
D\right)  =L_{SI}^{\ast}\left(  D\right)  $ (see Fig. \ref{Fig_Gauss}.%
\begin{figure*}[tbp] \centering
{\includegraphics[
trim=0.000000in 0.075158in 0.000000in 0.077715in,
height=2.8262in,
width=5.8496in
]%
{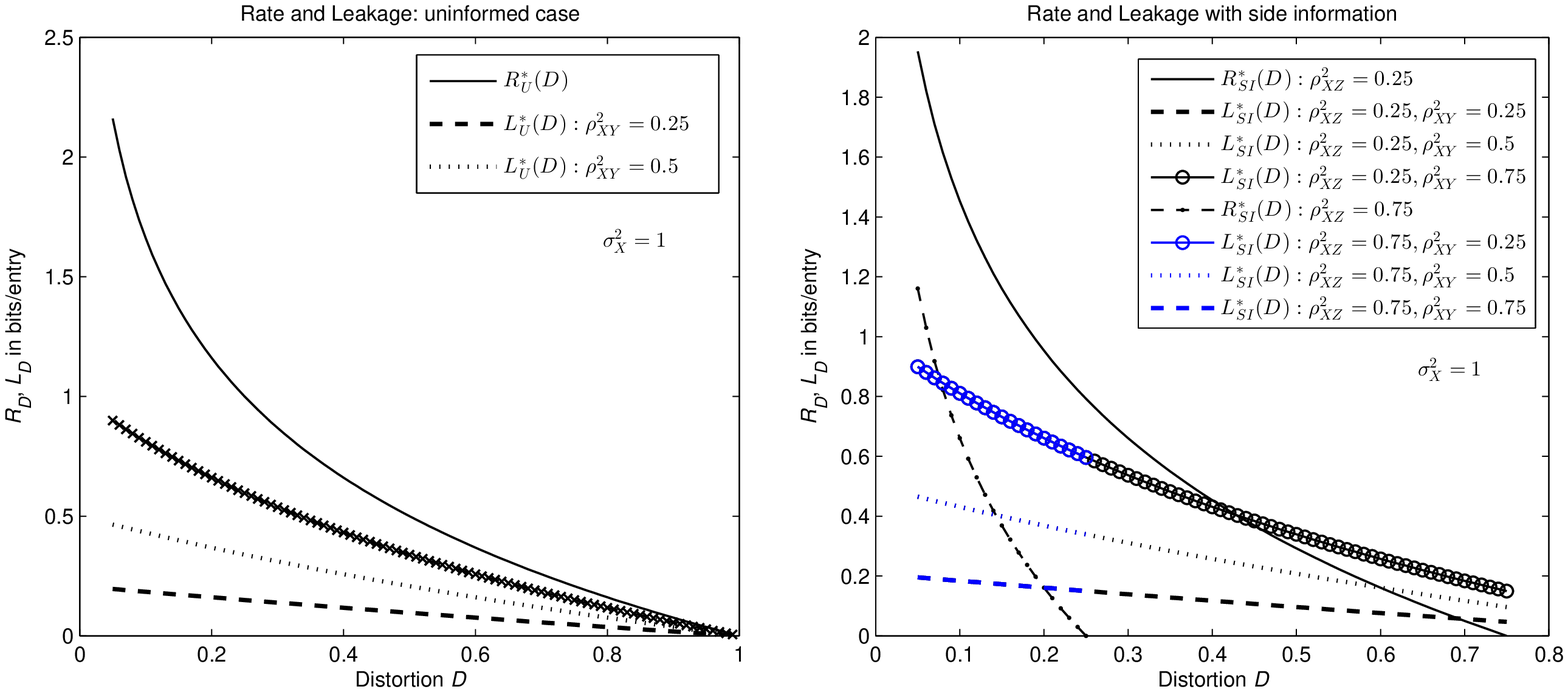}%
}
\caption{Plot of Rate and Leakage vs. D for Cases (i), (ii), and (iii).}\label{Fig_Gauss}%
\end{figure*}%

\emph{Interpretation}: The RDL and U-P tradeoffs for the Gaussian models
considered here reveal that the privacy-maximal code requires that the
reconstructed database is also Gaussian distributed. This in turn is a direct
result of the following fact: a Gaussian distribution has the maximal
(conditional and unconditional) entropy (uncertainty) for a fixed variance
\cite[Chap 8, Th. 8.6.5]{CTbook} (and hence, a fixed mean-squared distortion
between the input and output databases). Thus, if one wishes to preserve the
most uncertainty about the original input database from the output, the output
must also be Gaussian distributed, i.e., it suffices to add Gaussian noise,
since the sum of two Gaussians is a Gaussian. The power of our model and the
results are that not only can one find the privacy-optimal noise perturbation
for the Gaussian case but that practical applications such as medical
analytics that assume Gaussian-distributed data can still work on sanitized
data, albeit with modified parameter values.

In \cite{Kargupta}, it was noted that Gaussian noise is often the easiest to
filter and this observation may seem to be in conflict with our result -- if
the added noise can be filtered out, the privacy protection afforded by the
added noise can be reduced by the adversary. However, what \cite{Kargupta}
actually shows is that when the spectra of the noise and the data differ
significantly the noise can be filtered, thereby jeopardizing privacy
measures. For the i.i.d. source model (i.e., a source with no memory)
considered here, the i.i.d. Gaussian noise that is added to guarantee privacy
has the same flat power spectral density as the source, and thus, the
perturbed data cannot be distinguished from the added noise. In fact, the
quantization that underlies the information-theoretic sanitization mechanism
developed here is an irreversible process and one cannot obtain the original
data except for $D=0$ (i.e., the case of no sanitization). As a point of
comparison, we note that in a separate work on privacy of streaming data
(non-i.i.d time-series data modeled as a colored Gaussian process, i.e.\ data
that has non-flat spectrum), we have shown that the privacy-optimal noise
perturbation requires the spectrum of the added noise to be non-flat to match
that of the non-i.i.d. data \cite{LS_SG4}.

Our example also reveals how finding the optimal santization mechanism, i.e.,
the optimal mapping from the original public to the revealed attributes
depends both on the statistical model. In fact, it is for this reason that
adding Gaussian noise for any numerical database will not, in general, be
optimal unless the database statistics can be approximated by a Gaussian distribution.
\end{example}

\section{\label{Sec_V}Concluding Remarks}

The ability to achieve the desired level of privacy while guaranteeing a
minimal level of utility and vice-versa for a general data source is
paramount. Our work defines privacy and utility as fundamental characteristics
of data sources that may be in conflict and can be traded off. This is one of
the earliest attempts at systematically applying information theoretic
techniques to this problem. Using rate-distortion theory, we have developed a
U-P tradeoff region for i.i.d. data sources with known distribution.

We have presented a theoretical treatment of a universal (i.e. not dependent
on specific data features or adversarial assumptions) theory for privacy and
utility that addresses both numeric and categorical (non-numeric) data. We
have proposed a novel notion of privacy based on guarding existing uncertainty
about hidden data that is intuitive but also supported by rigorous theory.
Prior to our work there was no comparable model that applied to both data
types, so no side-by-side comparisons can be made across the board between
different approaches. The examples developed here are the first step towards
understanding practical approaches with precise guarantees. The next step
would be to pick specific sample domains (e.g., medical data, census data),
devise the appropriate statistical distributions and U-P metrics, set
desirable levels of privacy and utility parameters, and then analyze on test
data. These topics for future research however require the theoretical
framework proposed here as a crucial first step.

Several challenges remain in quantifying utility-privacy tradeoffs for more
general sources. For example, our model needs to be generalized for non-i.i.d.
data sources, sources with unknown distributions, and sources lacking strong
structural properties (such as Web searches). Results from rate-distortion
theory for sources-with-memory and universal lossy compression may help
address these challenges. Farther afield, our privacy guarantee is an average
metric based on Shannon entropy which may be inadequate for some applications
where strong anonymity guarantees are required for every individual in a
database (such as an HIV database). Finally, we have recently extended this
framework to privacy applications with time-series sources \cite{LS_SG4} and
organizational data disclosure \cite{LS_SG1}.%

\appendix
{}

\subsection{\label{App1}Proofs of Theorems \ref{TheoremS_informed} and
\ref{Theorem_Inf}}

\subsubsection{Statistically Informed Case: Proof of Theorem
\ref{TheoremS_informed}}

\textit{Converse}: We now formally develop lower and upper bounds on the rate
and equivocation, respectively, that is achievable for the statistically
informed encoder case. We show that given a $(n,2^{n\left(  R+\epsilon\right)
},D+\epsilon,E-\epsilon)$ code there exists a $p(x_{\mathcal{K}_{r}%
},x_{\mathcal{K}_{h}},z)p\left(  u|x_{\mathcal{K}_{r}},x_{\mathcal{K}_{h}%
}\right)  $ such that the rate and equivocation of the system are bounded as follows:%

\begin{align}
R+\epsilon &  \geq\frac{1}{n}\log M\geq\frac{1}{n}H(J)\geq\frac{1}{n}I\left(
J;X_{\mathcal{K}}^{n}|Z^{n}\right) \nonumber\\
&  =\frac{1}{n}\left\{  H(X_{\mathcal{K}}^{n}|Z^{n})-H(X_{\mathcal{K}}%
^{n}|JZ^{n})\right\} \\
&  =\frac{1}{n}%
{\textstyle\sum\limits_{i=1}^{n}}
H(X_{\mathcal{K},i}|Z_{i})\label{ApWZIID}\\
&  \text{ \ \ \ }-\frac{1}{n}%
{\textstyle\sum\limits_{i=1}^{n}}
H\left(  X_{\mathcal{K},i}|X_{\mathcal{K}}^{i-1}Z_{i}\left(  JZ^{i-1}%
Z_{i+1}^{n}\right)  \right) \nonumber
\end{align}%
\begin{align}
&  \geq\frac{1}{n}%
{\textstyle\sum\limits_{i=1}^{n}}
H(X_{\mathcal{K},i}|Z_{i})-\frac{1}{n}%
{\textstyle\sum\limits_{i=1}^{n}}
H\left(  X_{\mathcal{K},i}|Z_{i}U_{i}\right) \label{ApWZMI}\\
&  =\frac{1}{n}%
{\textstyle\sum\limits_{i=1}^{n}}
R_{SI}\left(  D_{i}\right) \label{ROBSI7}\\
&  \geq R_{SI}\left(  D\right)  \label{ROBSI8}%
\end{align}
where $X_{\left(  \cdot\right)  }^{i-1}=[X_{\left(  \cdot\right)  ,1}$
$X_{\left(  \cdot\right)  ,2}$ $\ldots$ $X_{\left(  \cdot\right)  ,i-1}]$,
$i\geq1$, (\ref{ApWZIID}) follows from the assumption of an i.i.d. source,
(\ref{ApWZMI}) from the fact that conditioning does not increase entropy and
by setting $U_{i}\equiv\left(  JZ^{i-1}Z_{i+1}^{n}\right)  $ such that
$U_{i}-X_{\mathcal{K}}-Z_{i}$ forms a Markov chain for all $i,$ and $\hat
{X}_{\mathcal{K}_{r},i}=g_{i}\left(  J,Z^{n}\right)  =f_{i}\left(  U_{i}%
,Z_{i}\right)  $ for some $g_{i}$ and $f_{i}$, (\ref{ROBSI7}) from definition
(\ref{R_WZ}) for
\[
D_{i}\equiv\mathbb{E}\left[  d\left(  X_{\mathcal{K},i},\hat{X}_{\mathcal{K}%
,i}\right)  \right]  ,\text{ and }E_{SI.i}\equiv H(Y_{i}|U_{i}Z_{i}),
\]
and (\ref{ROBSI8}) from the convexity of the function $R_{SI}\left(  D\right)
$ defined in (\ref{R_WZ}) (see \cite[Chap. 10]{CTbook}, \cite{Wyner_Ziv}).

For the same $(n,2^{n\left(  R+\epsilon\right)  },D,E-\epsilon)$ code
considered, we can upper bound the achievable equivocation as
\begin{align}
E-\epsilon &  \leq\frac{1}{n}H\left(  X_{\mathcal{K}_{h}}^{n}|JZ^{n}\right)
\nonumber\\
&  =\frac{1}{n}%
{\textstyle\sum\limits_{i=1}^{n}}
H\left(  X_{\mathcal{K}_{h},i}|X_{\mathcal{K}_{h}}^{i-1}Z_{i}\left(
JZ^{i-1}Z_{i+1}^{n}\right)  \right) \nonumber\\
&  \leq\frac{1}{n}%
{\textstyle\sum\limits_{i=1}^{n}}
H\left(  X_{\mathcal{K}_{h},i}|Z_{i}U_{i}\right) \label{ESI_3}\\
&  =\frac{1}{n}%
{\textstyle\sum\limits_{i=1}^{n}}
E_{SI}\left(  D_{i}\right) \label{ESI_4}\\
&  \leq E_{SI}\left(  D\right)  \label{ESI_5}%
\end{align}
where (\ref{ESI_4}) follows from (\ref{E_WZ}) and (\ref{ESI_5}) follows from
the concavity of the equivocation (logarithm) function $E_{SI}$.

\begin{remark}
If the private variables $X_{\mathcal{K}_{h}}^{n}$ are not directly used in
encoding, i.e., $X_{\mathcal{K}_{h}}^{n}-X_{\mathcal{K}_{r}}^{n}-U^{n}$ form a
Markov chain, then from the i.i.d. assumption of the source and the resulting
encoding, the Markov chain $X_{\mathcal{K}_{h},i}-X_{\mathcal{K}_{r},i}-U_{i}$
holds for all $i=1,2,\ldots,n$.
\end{remark}

\textit{Achievability}: We briefly summarize the quantize-and-bin coding
scheme for the statistically informed encoder case. Consider an input
distribution $p(u,x_{\mathcal{K}},z)$:
\[
p(u,x_{\mathcal{K}},z)=p(u,x_{\mathcal{K}})p(z|x_{\mathcal{K}}),
\]
i.e., $U-X_{\mathcal{K}}-Z$ forms a Markov chain. Fix $p\left(
u|x_{\mathcal{K}}\right)  $. First generate $M=2^{n\left(  I(U;X_{\mathcal{K}%
})+\epsilon\right)  }$, $U^{n}\left(  w\right)  $ databases, $w=1,2,\ldots,M$,
i.i.d. according to $p\left(  u\right)  $. Let $W$ denote the random variable
for the index $w$. Next, for ease of notation, denote the following:
\[
S=2^{nI(X_{\mathcal{K}};U)},R=2^{nI(X_{\mathcal{K}};U|Z)},T=2^{nI(U;Z)}.
\]
The encoder bins the $u^{n}(w)$ sequences into $R$ bins as follows:
\[
J(u^{n}(w))=k,\text{if }w\in\lbrack(k-1)T+1,kT].
\]

Upon observing a source sequence $x_{\mathcal{K}}^{n},$ the encoder searches
for a $u^{n}\left(  w\right)  $ sequence such that $\left(  x_{\mathcal{K}%
}^{n},u^{n}\left(  w\right)  \right)  \in\mathcal{T}_{X_{\mathcal{K}}U}\left(
n,\epsilon\right)  $ (the choice of $M$ ensures that there exists at least one
such $w$). The encoder sends $J\left(  w\right)  $ where $J\left(  w\right)  $
is the bin index of $u^{n}\left(  w\right)  $ sequence sent at a rate
$R=I(X_{\mathcal{K}};U|Z)+\epsilon$.

This encoding scheme implies the decodability of $U^{n}$ sequence as follows:
upon receiving the bin index $J(u^{n}(w))=j$, the uncertainty at the decoder
about $u^{n}(w)$ is reduced. In particular, having the bin index $j$, it knows
that there are only $2^{nI(U;Z)}$ possible $u^{n}$ sequences that could have
resulted in the bin index $j$. It then uses joint typical decoding using
$Z^{n}$ to decode the correct $u^{n}$ sequence (the probability of decoding
error goes to zero as $n\rightarrow\infty$ by standard arguments as in the
channel coding theorem). This implies that using Fano's inequality, the
decoder having access to $(J,Z^{n})$ can correctly $W$, and hence, decode
$U^{n}\left(  W\right)  ,$ with high probability, i.e.,
\begin{equation}
\frac{1}{n}H(W|J,Z^{n})=\frac{1}{n}H(U^{n}(W)|J,Z^{n})\leq\delta(n),
\label{Fano_SI}%
\end{equation}
where $\delta(n)\rightarrow0$ as $n\rightarrow\infty$.

\subsubsection{Proof of Equivocation}

For the quantize-and-bin scheme presented above, we will show that
\[
\lim_{n\rightarrow\infty}\frac{1}{n}H(X_{\mathcal{K}_{h}}^{n}|J,Z^{n})\geq
H(X_{\mathcal{K}_{h}}|U,Z)-\epsilon,
\]
which is equivalent to showing that
\[
\lim_{n\rightarrow\infty}\frac{1}{n}I(X_{\mathcal{K}_{h}}^{n};J,Z^{n})\leq
I(X_{\mathcal{K}_{h}};U,Z)+\epsilon.
\]

Our proof is based on the fact that for the chosen quantize-and-bin coding
scheme, at the decoder given the bin index and side information, the
uncertainty of the quantized sequences $U^{n}$ approaches zero for large $n$
as shown in (\ref{Fano_SI}).

Consider the term $I(X_{\mathcal{K}_{h}}^{n};J,U^{n},Z^{n})$ which can be
written as
\begin{subequations}
\label{EqAchSI}%
\begin{align}
&  I(X_{\mathcal{K}_{h}}^{n};J,Z^{n})+I(X_{\mathcal{K}_{h}}^{n};U^{n}%
|J,Z^{n})\\
&  =I(X_{\mathcal{K}_{h}}^{n};J,Z^{n})\label{EqBnd_2}\\
&  =I(X_{\mathcal{K}_{h}}^{n};U^{n},Z^{n})+I(X_{\mathcal{K}_{h}}^{n}%
;J|U^{n},Z^{n})\label{EqBnd_3}\\
&  \leq I(X_{\mathcal{K}_{h}}^{n};U^{n},Z^{n})\label{EqBnd_4}\\
&  =nH(X_{\mathcal{K}_{h}})-H\left(  X_{\mathcal{K}_{h}}^{n}|U^{n}%
,Z^{n}\right) \label{EqBnd_4a}\\
&  \leq n\left(  I(X_{\mathcal{K}_{h}};U,Z)+\delta\left(  n\right)  \right)
\label{EqBnd_5}\\
&  \leq n\left(  I(X_{\mathcal{K}_{h}};U,Z)+\epsilon\right)  \label{EqBnd_6}%
\end{align}
where (\ref{EqBnd_2}) follows from (\ref{Fano_SI}), (\ref{EqBnd_3}) follows
from (\ref{Fano_SI}) and the fact that the mutual information is strictly
non-negative, (\ref{EqBnd_4}) follows from the fact that there is no
uncertainty in bin index $J(W)$ given $U^{n}\left(  W\right)  $,
(\ref{EqBnd_4a}) follows from the i.i.d. assumption on the source and side
information statistics, (\ref{EqBnd_5}) is proved in \ref{Appbound} below such
that $\delta\left(  n\right)  \rightarrow0$ as $n\rightarrow\infty$, and
finally (\ref{EqBnd_6}) follows from choosing $\epsilon\geq\delta\left(
n\right)  $ that determines the size $M=2^{n\left(  R+\epsilon\right)  }$ of
the codebook arbitrarily small as $n\rightarrow\infty$.

\subsubsection{Informed Encoder Case: Proof of Theorem \ref{Theorem_Inf}}

\textit{Converse}: We now formally develop lower and upper bounds on the rate
and equivocation, respectively, that is achievable for the informed encoder
case. The converse for the rate mirrors standard converse and we clarify the
steps briefly. We show that given a $(n,2^{n\left(  R+\epsilon\right)
},D+\epsilon,E-\epsilon)$ code there exists a $p(x_{\mathcal{K}_{r}%
},x_{\mathcal{K}_{h}},z)p\left(  \hat{x}_{\mathcal{K}_{r}}|x_{\mathcal{K}_{r}%
},x_{\mathcal{K}_{h}},z\right)  $ such that the rate and equivocation of the
system are bounded as follows:%

\end{subequations}
\begin{align}
R+\epsilon &  \geq\frac{1}{n}H(J)\geq\frac{1}{n}I\left(  J;X_{\mathcal{K}}%
^{n},Z^{n}\right)  \geq\frac{1}{n}I(X_{\mathcal{K}}^{n};J|Z^{n})\nonumber\\
&  \geq\frac{1}{n}%
{\textstyle\sum\limits_{i=1}^{n}}
H(X_{\mathcal{K},i}|Z_{i})-\frac{1}{n}%
{\textstyle\sum\limits_{i=1}^{n}}
H\left(  X_{\mathcal{K},i}|JZ^{n}\hat{X}_{\mathcal{K}_{r}}^{n}\right)
\nonumber\\
&  \geq\frac{1}{n}%
{\textstyle\sum\limits_{i=1}^{n}}
H(X_{\mathcal{K},i}|Z_{i})-\frac{1}{n}%
{\textstyle\sum\limits_{i=1}^{n}}
H\left(  X_{\mathcal{K},i}|Z_{i}\hat{X}_{\mathcal{K}_{r},i}\right) \nonumber\\
&  =\frac{1}{n}%
{\textstyle\sum\limits_{i=1}^{n}}
R_{SI}\left(  D_{i}\right) \label{ApInf2}\\
&  \geq R_{SI}\left(  D\right)  \label{ApInf3}%
\end{align}
where (\ref{ApInf3}) follows from the convexity of the function $R_{I}\left(
D\right)  $ defined in (\ref{R_WZ}) \cite[Chap. 10]{CTbook} for
\begin{subequations}
\begin{align}
D_{i}  &  \equiv\mathbb{E}\left[  d\left(  X_{\mathcal{K},i},\hat
{X}_{\mathcal{K},i}\right)  \right]  ,\text{ and}\\
E_{I.i}  &  \equiv H(Y_{i}|\hat{X}_{\mathcal{K},i}). \label{E_Inf}%
\end{align}
For the same $(n,2^{n\left(  R+\epsilon\right)  },D,E-\epsilon)$ code
considered, we can upper bound the achievable equivocation as
\end{subequations}
\begin{align}
E-\epsilon &  \leq\frac{1}{n}H\left(  X_{\mathcal{K}_{h}}^{n}|JZ^{n}\right)
\nonumber\\
&  =\frac{1}{n}%
{\textstyle\sum\limits_{i=1}^{n}}
H\left(  X_{\mathcal{K}_{h},i}|X_{\mathcal{K}_{h}}^{i-1}Z^{n}J\hat
{X}_{\mathcal{K}_{r}}^{n}\right) \label{InfE1}\\
&  \leq\frac{1}{n}%
{\textstyle\sum\limits_{i=1}^{n}}
H\left(  X_{\mathcal{K}_{h},i}|Z_{i}\hat{X}_{\mathcal{K}_{r},i}\right)
\label{InfE2}\\
&  =\frac{1}{n}%
{\textstyle\sum\limits_{i=1}^{n}}
E_{I}\left(  D_{i}\right) \label{InfE3}\\
&  \leq E_{I}\left(  D\right)  \label{InfE4}%
\end{align}
where (\ref{InfE1}) follows from the fact that the reconstructed database
$\hat{X}_{\mathcal{K}_{r}}^{n}$ is a function of the $J$ and $Z^{n},$
(\ref{InfE3}) follows from the fact that conditioning does not increase
entropy, (\ref{InfE3}) follows from (\ref{E_Inf}), and (\ref{ESI_5}) follows
from the concavity of the equivocation (logarithm) function $E_{I}$.

\begin{remark}
If the hidden variables $X_{\mathcal{K}_{h}}^{n}$ are not directly used in
encoding, i.e., $X_{\mathcal{K}_{h}}^{n}-X_{\mathcal{K}_{r}}^{n}-\hat
{X}_{\mathcal{K}_{r}}^{n}$ form a Markov chain, then from the i.i.d.
assumption of the source and the resulting encoding, the Markov chain
$X_{\mathcal{K}_{h},i}-X_{\mathcal{K}_{r},i}-\hat{X}_{\mathcal{K}_{r}i}$ holds
for all $i=1,2,\ldots,n$.
\end{remark}

\textit{Achievability}: We briefly summarize the quantize-and-bin coding
scheme for the informed encoder case. The encoding mirrors that for the
statistically informed case and in the interest of space only the differences
are highlighted below. The primary difference is that the database encoder now
encodes both $\left(  X_{\mathcal{K}},Z\right)  $ such that the input
distribution $p(x_{\mathcal{K}},\hat{x}_{\mathcal{K}_{r}},z)$ is%
\[
p(x_{\mathcal{K}},\hat{x}_{\mathcal{K}_{r}},z)=p(z,x_{\mathcal{K}})p(\hat
{x}_{\mathcal{K}_{r}}|x_{\mathcal{K}},z).
\]
i.e., $\hat{X}_{\mathcal{K}_{r}}$ is a function of both $X_{\mathcal{K}}$ and
$Z$. This distribution is now used to generate $M=2^{n\left(  I(\hat
{X}_{\mathcal{K}_{r}};X_{\mathcal{K}}Z)+\epsilon\right)  }$, $\hat
{X}_{\mathcal{K}_{r}}^{n}\left(  w\right)  $ sequences as before which are
first quantized and then binned at a rate $R=2^{nI(X_{\mathcal{K}};\hat
{X}_{\mathcal{K}_{r}}|Z)}.$ Decoding follows analogously to the previous case,
i.e., the decoder uses $Z^{n}$ and the bin index $J$ to decode the correct
$\hat{x}_{\mathcal{K}_{r}}^{n}$ sequence (the probability of decoding error
goes to zero as $n\rightarrow\infty$ by standard arguments as in the channel
coding theorem). This implies that using Fano's inequality, the decoder having
access to $(J,Z^{n})$ can correctly decode $W$, and hence, $\hat
{X}_{\mathcal{K}_{r}}^{n}\left(  W\right)  ,$ with high probability, i.e.,
\begin{equation}
\frac{1}{n}H(W|J,Z^{n})=\frac{1}{n}H(\hat{X}_{\mathcal{K}_{r}}^{n}%
(W)|J,Z^{n})\leq\epsilon(n), \label{Fano_Inf}%
\end{equation}
where $\epsilon(n)\rightarrow0$ as $n\rightarrow\infty$.

\textit{Proof of equivocation: }For the quantize-and-bin scheme presented
above, we need to show that
\[
\lim_{n\rightarrow\infty}\frac{1}{n}H(X_{\mathcal{K}_{h}}^{n}|J,Z^{n})\geq
H(X_{\mathcal{K}_{h}}|\hat{X}_{\mathcal{K}_{r}},Z)-\epsilon.
\]
Our proof is based on the fact that for the chosen quantize-and-bin coding
scheme, at the decoder given the bin index $J$ and side information $Z^{n}$,
the uncertainty of the quantized sequences $\hat{X}_{\mathcal{K}_{r}}$
approaches zero for large $n$ as shown in (\ref{Fano_Inf}). The proof is the
same as (\ref{EqAchSI}) with $U=\hat{X}_{\mathcal{K}_{r}}$ along with
(\ref{Fano_Inf}) and is omitted for brevity.

\subsection{\label{Appbound}Proof of (\ref{EqBnd_5})}

Here, we prove the following inequality:
\[
H(X_{\mathcal{K}_{h}}^{n}|U^{n},Z^{n})\leq n(H(X_{\mathcal{K}_{h}%
}|U,Z)+\epsilon(n)).
\]
For ease of exposition, let $Y^{n}\equiv X_{\mathcal{K}_{h}}^{n}$ such that
$H(X_{\mathcal{K}_{h}}^{n}|U^{n},Z^{n})=H(Y^{n}|U^{n},Z^{n})$ can be expanded
and bounded as%
\begin{align*}
&  =\sum_{(\mathbf{u},\mathbf{z})}p(\mathbf{u},\mathbf{z})H(Y^{n}%
|U^{n}=\mathbf{u},Z^{n}=\mathbf{z})\\
&  =\sum_{(\mathbf{u},\mathbf{z})\in\mathcal{T}_{UZ}}p(\mathbf{u}%
,\mathbf{z})H(Y^{n}|U^{n}=\mathbf{u},Z^{n}=\mathbf{z})\\
&  \quad+\sum_{(\mathbf{u},\mathbf{z})\notin\mathcal{T}_{UZ}}p(\mathbf{u}%
,\mathbf{z})H(Y^{n}|U^{n}=\mathbf{u},Z^{n}=\mathbf{z})
\end{align*}
\begin{align*}
&  \leq\sum_{(\mathbf{u},\mathbf{z})\in\mathcal{T}_{UZ}}p(\mathbf{u}%
,\mathbf{z})H(Y^{n}|U^{n}=\mathbf{u},Z^{n}=\mathbf{z})\\
&  \quad+\sum_{(\mathbf{u},\mathbf{z})\notin\mathcal{T}_{UZ}}p(\mathbf{u}%
,\mathbf{z})nH(Y)\\
&  \leq\sum_{(\mathbf{u},\mathbf{z})\in\mathcal{T}_{UZ}}p(\mathbf{u}%
,\mathbf{z})(Y^{n}|U^{n}=\mathbf{u},Z^{n}=\mathbf{z})\\
&  \quad+nH(Y)\delta(n)
\end{align*}%
\begin{align*}
&  =\sum_{(\mathbf{u},\mathbf{z})\in\mathcal{T}_{UZ}}p(\mathbf{u}%
,\mathbf{z})\Bigg[-\sum_{\mathbf{y}}p(\mathbf{y}|\mathbf{u},\mathbf{z}%
)\log(p(\mathbf{y}|\mathbf{u},\mathbf{z}))\Bigg]\\
&  \quad+nH(Y)\delta(n)\\
&  =\sum_{(\mathbf{u},\mathbf{z})\in\mathcal{T}_{UZ}}p(\mathbf{u}%
,\mathbf{z})\Bigg[-\sum_{\mathbf{y}\in\mathcal{T}_{Y|\mathbf{u},\mathbf{z}}%
}p(\mathbf{y}|\mathbf{u},\mathbf{z})\log(p(\mathbf{y}|\mathbf{u}%
,\mathbf{z}))\\
&  \quad\quad-\sum_{\mathbf{y}\notin\mathcal{T}_{Y|\mathbf{u},\mathbf{z}}%
}p(\mathbf{y}|\mathbf{u},\mathbf{z})\log(p(\mathbf{y}|\mathbf{u}%
,\mathbf{z}))\Bigg]+nH(Y)\delta(n)
\end{align*}%
\begin{align*}
&  \leq\sum_{(\mathbf{u},\mathbf{z})\in\mathcal{T}_{UZ}}p(\mathbf{u}%
,\mathbf{z})\Bigg[-\sum_{\mathbf{y}\in\mathcal{T}_{Y|\mathbf{u},\mathbf{z}}%
}p(\mathbf{y}|\mathbf{u},\mathbf{z})\log(p(\mathbf{y}|\mathbf{u}%
,\mathbf{z}))\Bigg]\\
&  \quad\quad+nH(X)\delta(n)+\epsilon(n)\\
&  \leq n(H(Y|U,Z)+2\epsilon(n)+H(Y)\delta(n))\\
&  =n(H(Y|U,Z)+\zeta(n)),
\end{align*}
where $\zeta(n)\rightarrow0$ as $n\rightarrow\infty$.

\bibliographystyle{IEEEtran}
\bibliography{DB_refs}

\end{document}